\documentclass[10pt,journal]{IEEEtran}

\usepackage{graphicx}
\usepackage{amsmath,amssymb} 
\usepackage{color}
\usepackage{overpic}
\usepackage[ruled]{algorithm2e}
\usepackage{soul}
\usepackage{cite}
\usepackage{xcolor,colortbl}
\usepackage{multirow}
\usepackage{subfigure}
\usepackage{url}
\usepackage{booktabs}

\newcommand{\tabincell}[2]{\begin{tabular}{@{}#1@{}}#2\end{tabular}}

\newcommand{\wh}[1]{{\color{black}#1}}
\newcommand{\wht}[1]{{\color{black}#1}}
\newcommand{\whtt}[1]{{\color{black}#1}}

\begin{document}

\title{\wht{A Comprehensive Benchmark for Single Image Compression Artifacts Reduction}}
\author{
Jiaying Liu, \textit{Senior Member, IEEE}, Dong Liu, \textit{Senior Member, IEEE}, \\ Wenhan Yang, \textit{Member, IEEE}, Sifeng Xia, Xiaoshuai Zhang, Yuanying Dai
}
%\author{
%	A, \textit{Senior Member, IEEE}, A, \textit{Senior Member, IEEE}, \\ C, \textit{Member, IEEE}, D
%}
\markboth{IEEE Transactions on Image Processing, Draft}%
{Shell \MakeLowercase{\textit{et al.}}}

\maketitle

\begin{abstract}
We present a comprehensive study and evaluation of existing single image compression artifacts removal algorithms, using a new 4K resolution benchmark including diversified foreground objects and background scenes with rich structures, called Large-scale Ideal Ultra high definition 4K (LIU4K) \wh{benchmark}.
Compression artifacts removal, as a common post-processing technique, aims at alleviating undesirable artifacts such as \wh{blockiness}, ringing, and banding caused by quantization and approximation in the compression process.
In this work, a systematic listing of the reviewed methods is presented based on their basic models (handcrafted models and deep networks).
The main contributions and novelties of these methods are highlighted, and
the main development directions, including architectures, multi-domain sources, signal structures, and new targeted units, are summarized.
Furthermore, based on a unified deep learning configuration (\textit{i.e.} same training data, loss function, optimization algorithm, \textit{etc.}), we  evaluate recent deep learning-based methods based on diversified evaluation measures.
The experimental results show the state-of-the-art performance comparison of existing methods based on both full-reference, non-reference and task-driven metrics.
%At last, we also explore novel constraints and training skills that may benefit a series of compression artifacts removal methods.
Our survey would give a comprehensive reference source for future research on single image compression artifacts removal and inspire new directions of the related fields.
\end{abstract}

\begin{IEEEkeywords}
	Compression artifacts removal, benchmark, side information, loop filter, deep learning
\end{IEEEkeywords}

\renewcommand{\baselinestretch}{0.97}

\section{Introduction}
\label{sec:intro}

\IEEEPARstart{L}ossy compression, such as JPEG~\cite{JPEG_Standard}, HEVC~(High Efficiency Video Coding)~\cite{HEVC}, Advanced Video Coding~(AVC)~\cite{AVC}, has been widely used in image and video codecs to reduce information redundancy in transmission and storage process to save bandwidth and resource.
Based on human visual properties, the codecs make use of redundancies in spatial, temporal and transform domains to provide approximations of encoded content compactly.
They effectively reduce the bit-rate cost but inevitably lead to unsatisfactory visual artifacts, \textit{e.g.} blockiness, ringing and banding. These artifacts are derived from the high-frequency detail loss in the quantization process and the discontinuities caused by block-wise batch processing. These artifacts not only degrade user visual experience, but also are detrimental for the successive image processing and computer vision tasks.

In our work, we focus on the degradation of compressed images. The degradation configurations of two codecs are considered: JPEG and HEVC.
Most modern codecs first divide the whole image into blocks, which sometimes have a fixed size, \textit{e.g.} JPEG, while others have different sizes, \textit{e.g.} HEVC. Then, transformations, \textit{e.g.} discrete cosine transformation (DCT) and discrete sine transformation (DST) \textit{etc.}, follow to convert each block into transformed coefficients with more compact energy and sparser distributions than those in the spatial domain. After that, quantization is applied on the transformed coefficients, based on the pre-defined quantization steps, to remove the signal components that have less significant influence on \wh{the} human visual system. The quantization intervals are usually much larger on high-frequency components than those on low-frequency components, because human visual system is less capable of distinguishing high frequency components. 
Noting that the quantization step is the main reason causing artifacts. After quantization, the boundaries between blocks become discontinuous. Thus,  \textit{blocking artifacts} are generated.
\textit{Blurring} is caused by the loss of high-frequency components.
In regions including sharp edges, the \textit{ringing} artifacts become visible. 
When the quantization step becomes larger, the reconstructed images suffer from severe distortions. Noticeable \textit{banding effects} appear in smooth regions over the image.

Lots of efforts are dedicated to the restoration of compressed images. 
Early preliminary works~\cite{TV_jpeg,regresion_based} perform filtering along the boundaries to remove simple artifacts. 
After that, data-driven methods are then proposed to learn the inverse mapping of compression degradations to remove artifacts. 
These methods are proposed in two directions: 1) one for better inference models, \textit{e.g.}, sparse coding~\cite{learned_dict} and deep networks~\cite{ARCNN}; 2) the other for utilizing better priors and side information~\cite{d3,DDCN}. 
In recent years, the emergence of deep learning~\cite{ARCNN} largely improves the restoration capacity of the data-driven methods, with its excellent nonlinear modeling capacity. 
More advanced network architectures, \textit{e.g.} dense residual network~\cite{dense_residual}, are put forward and more strong side information, \textit{e.g.} partition mask~\cite{partition_mask}, is employed for compression artifacts removal.
Besides these two common factors, for learning-based approaches, the training configurations and protocols, including training data, losses, optimization approaches, data generation, and \wh{details of codecs} \textit{etc.}, also have large effects on the final performance. Thus, the related detail changes in these factors also contribute to the performance gains.

Despite their promising results, there are several neglected issues in previous methods. 
First, there is no \wh{unified} framework to understand and sort out all previous methods. A survey that compares and summarizes these methods with a simple and integrated view is needed.
Second, inconsistent experimental configurations and protocols are employed in different works. There is a lack of benchmarking efforts on state-of-the-art algorithms on a large-scale public dataset.
Last, previous datasets do not cover 4K resolution images, which in fact sets a barrier to compare the performance of different methods in recently more and more popular ultra high definition display devices.

\begin{figure*}[t]
	\centering
	\subfigure{
		\includegraphics[width=18cm]{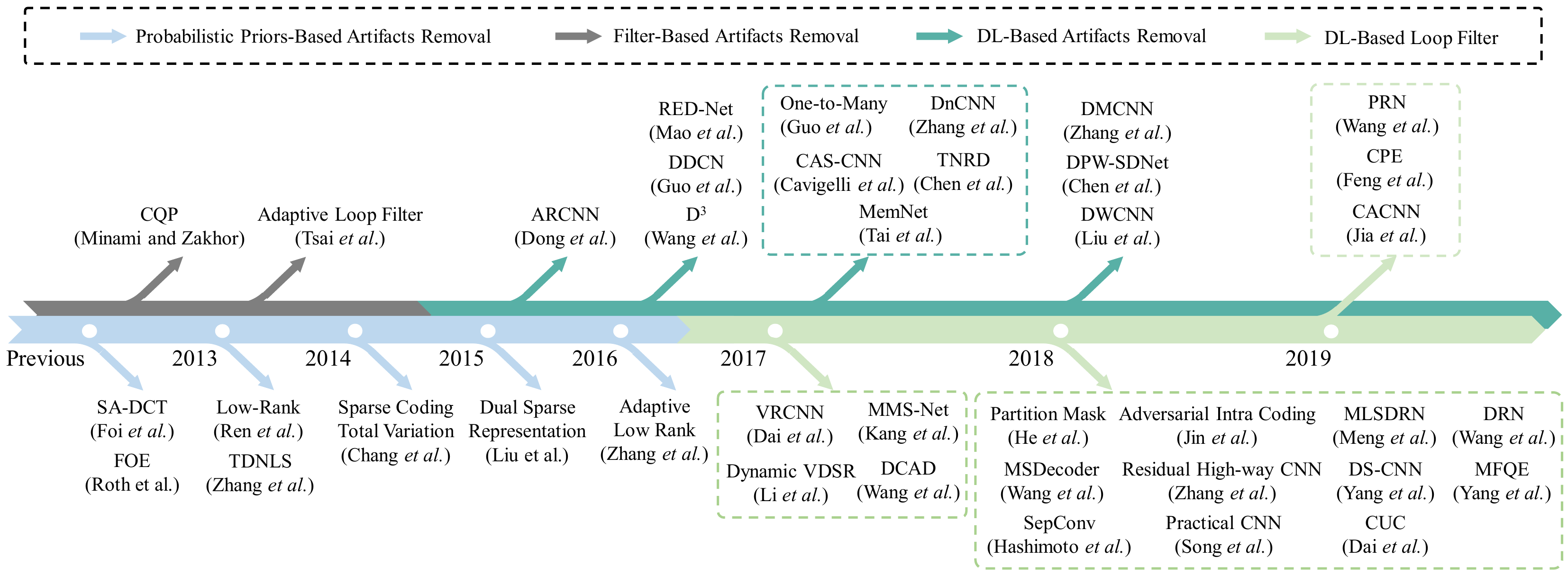}}
	\vspace{-6mm}
	\caption{
	Milestones of compressed image restoration methods, including filter based artifacts removal, probabilistic priors based artifacts removal, deep learning-based artifacts removal, and deep learning-based loop filter.
	The time period up to 2015 is dominated by handcrafted methods, including filter-based and probabilistic priors-based artifacts removal.
	The emergence of ARCNN~\cite{ARCNN} changes the development of this domain.
	A turning point is observed in 2015. After that, deep learning-based methods play a major role in the next several years. It is observed that, the years 2017 and 2018 welcome the blooming of deep learning-based artifacts removal and loop filters.
	}\vspace{-2mm}
	\label{fig:cv_task}
\end{figure*}

\begin{figure*}[t]
	\centering
	\subfigure{
		\includegraphics[width=18cm]{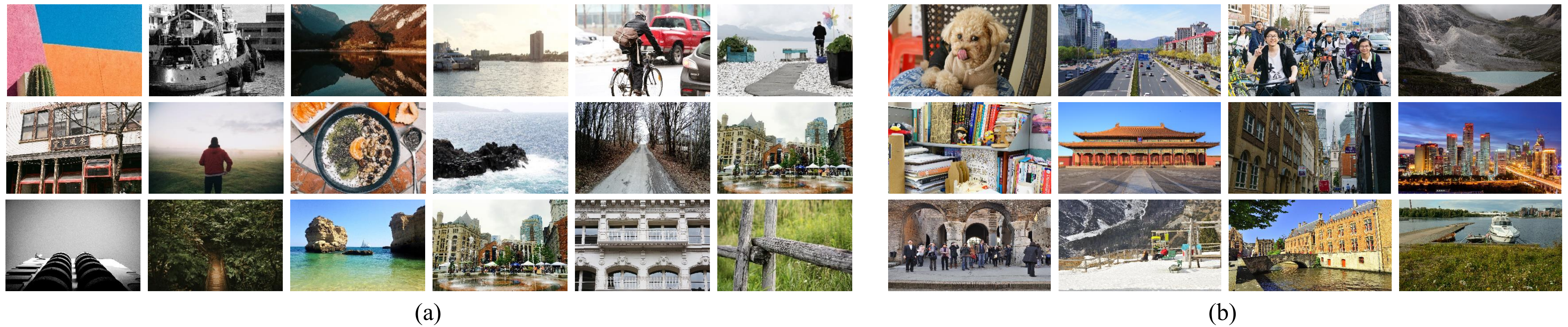}}
	\vspace{-6mm}
	\caption{Example images sampled from LIU4K. (a) Training set. (b) Testing set.} \vspace{-2mm}
	\label{fig:sample}
\end{figure*}

Our work paper is directly motivated to address the above
issues, and makes four-fold technical contributions:

\begin{itemize}	
	\item The first contribution of this paper is to provide a comprehensive survey on compression artifacts removal methods. Our survey provides a holistic view on most of the existing methods. 
	Particular \wh{emphasis} is placed on the deep learning-based single-image compression artifacts removal \wh{methods} as they offer the state-of-the-art performance and \wh{own much flexibility for further improvements}.
	\item We introduce a new single image compression artifact removal benchmark, called the Large-scale Ideal Ultra high definition 4K (LIU4K) dataset. It is the first dataset including 4K images as training and testing images serving for image restoration. It is also more large-scale than other existing datasets that include high definition images. Our LIU4K provides a better foundation to evaluate performance of different methods, especially in recent popular ultra high-resolution display devices.
	\item We conduct a systematic and extensive range of experiments to compare state-of-the-art methods quantitatively with diversified measures, on the new LIU4K dataset as well as previous commonly used datasets with a unified experimental setting, including the same training data, optimization method, and loss function \textit{et al}. The thorough evaluation and analysis show the performance and limitations of state-of-the-art methods. The new rich insights inspire new research directions.
	\item We also explore to generalize some constraints and training strategies from JPEG artifacts removal to the general compression artifacts removal. 
	Three strategies, including dense DCT transform constraints, mixed batches with the patches having different sizes, and gradually \wh{expanded} patch sizes are used in our experimental setting. 
	These strategies are also common to benefit future compression artifacts removal methods.
\end{itemize}
\vspace{-3mm}

%The rest of this paper is organized as follows. Section~\ref{sec:related_work} briefly reviews the related work. 
%Section~\ref{sec:retinex-net} presents our proposed Retinex-Net for robust low light image enhancement.
%Section~\ref{sec:retinex-net} illustrate our collected dataset.
%Experimental results and concluding remarks are presented in Sections~\ref{sec:exp} and~\ref{sec:conclusion}, respectively.

\section{A New Dataset for Restoration: LIU4K}
\label{sec:related_work}

\subsection{Previous Datasets}
We first review existing testing and training datasets: 
1) Testing: \textit{BSD100}, \textit{Kodak}, \textit{DIV2K-test}, \textit{Set5}, \textit{Set14}, \textit{Classic5}, and \textit{Twitter};
2) Training: \textit{BSD400}, \textit{DIV2K-train}, and \textit{Mini-ImageNet}.

\noindent \textbf{\textit{Kodak}}\footnote{http://r0k.us/graphics/kodak/}. It is a very representative dataset proposed in 1991 including 24 digital color images extracted from a wide range of films. After that, most image processing methods have been proposed, optimized, and evaluated based on this dataset. The resolution of images is 768$\times$512 or 512$\times$768. 

\noindent \textbf{\textit{BSD400}} and \textbf{\textit{BSD100}}. 
These two datasets are two parts of BSD500~\cite{bsd500}, which is originally designed for semantic segmentation.  
It covers a large variety of real-life scenes, with 200 training image, 200 validation image, and 100 testing images. The image resolution is 321$\times$481 or 481$\times$321.
For image restoration, we combine the training and validation sets of BSD500 as the training set of restoration and use its testing set for the restoration evaluation. 

\noindent \textbf{\textit{DIV2K}}~\cite{Timofte_2018_CVPR_Workshops}. 
There are 1,000 images whose resolution is 2K. 
\wh{It includes 800 images for training, 100 images for validation, and 100 images for testing.}
\wh{Sizes of the images are around 2000$\times$1000 or 1000$\times$2000.}
The max length of the height and width of an image is 2,040 and the other one is greater than 1,000.
It is a milestone dataset for image super-resolution and supports the NTIRE Challenge\footnote{http://www.vision.ee.ethz.ch/ntire17/} which uncovers preludes of challenges in low-level image enhancement. 

\noindent \textbf{\textit{Set5}}~\cite{Set5} and \textbf{\textit{Set14}}~\cite{Set14}.
They are two effective small-scale datasets to evaluate image restoration quality and usually provide consistent evaluation results to those on large-scale datasets. The resolution of \textit{Set5} is less than 500 $\times$ 500. The image size of \textit{Set14} is greater than $250\times250$ and less than $500\times500$. 

\noindent \textbf{\textit{Classic5}}~\cite{sa_dct}. \textit{Classic5} dataset includes five represented images used for evaluating compressed image restoration. Their resolutions are $512\times512$. 

\noindent \textbf{\textit{Twitter}}~\cite{ARCNN}. 40 images compressed by the Twitter platform whose sizes vary from 3,264 $\times$ 2,448 to 600 $\times$ 450. The included artifacts are highly complex because the compression process includes the rescaling operation. 

\noindent \textbf{\textit{Mini-ImageNet}}~\cite{Ledig_2017_CVPR}. The dataset used to train SRGAN in~\cite{Ledig_2017_CVPR} including 300,000 images sampled from ImageNet.
The smallest size is less than 50$\times$50, and the maximum size is larger than 4,000$\times$3,000. Although it might lead to superior performance of the restoration models, it is very time and resource-consuming to train with it.

\begin{table}[t]
	\scriptsize
	\caption{The summary of different datasets for compression artifact removal.}
	\label{tab:data_summary}
	\centering
	\begin{tabular}{ccccc}
		\hline
		Dataset & Number & Resolution & Train/Test & Features \\
		\hline
		{Kodak} & 24 & 768$\times$512 & Test & Earliest Milestone\\
		\hline
		{Set5} & 5 & 500 $\times$ 500 & Test & Small and Effective \\
		\hline
		{Set14} & 14 & \tabincell{c}{250 $\times$ 250 - \\ 500 $\times$ 500} & Test & Small, Effective \\
		\hline
		{Classic5} & 5 &  $512\times512$ & Test & Small, Effective \\
		\hline
		\tabincell{c}{BSD500} & \tabincell{c}{200/200/100} & $321\times481$  & \tabincell{c}{Train / Test \\ Validation } &
		\tabincell{c}{Middle Scale with \\ Abundant Texture} \\
		\hline
		\tabincell{c}{Mini- \\ ImageNet} & 300,000 & \tabincell{c}{50$\times$50-\\ 4,000$\times$3,000} & Train & Very Large Scale \\
		\hline
		{Twitter} & 40 & 600 $\times$ 450 & Test & Complex Degradation \\
		\hline
		{DIV2K} & \tabincell{c}{800/100/100} & 2,040 $\times$ 1,000 & \tabincell{c}{Train \& Test \\ Validation} & \tabincell{c}{Large-Scale with \\ 2K Images} \\
		\hline
		{LIU4K} & 1,500/200 & 3,264 $\times$ 2,448 & Train\&Test & \tabincell{c}{Large-Scale with \\ 4K Images} \\						
		\hline
	\end{tabular}
	\vspace{-6mm}
\end{table}

\begin{table*}[t]
	\caption{The statistical comparisons of different testing sets. 
	\wht{The numbers in brackets denote the variance. 
	The best results are denoted in bold.}}
	\vspace{-2mm}
	\label{tab:quality_metrics}
	\centering
	\begin{tabular}{cccccccc}
		\hline
		Dataset   & BPP    & PPI (10$^5$)  & Number & ENTROPY & BRISQUE & ENIQA ($10^{-4}$) & NIQE   \\
		\hline
		Set5      & 0.52 (0.004)  & 1.13  & 5      & 6.84  (0.548)   & 33.58 (47.82)   & 0.1393 (74.75)  & 4.79 (4.62)  \\
		Classic5  & 0.65 (0.004)  & 2.62  & 5      & 7.37  (0.037)   & 22.99 (146.58)  & 0.3696 (3.46)  & 5.08 (1.62) \\
		Kodak     & 0.56 (0.006)  & 3.93  & 24     & 6.93  (0.175)   & 6.22  (17.32)   & 0.0202 (6.35)  & 2.95 (0.21) \\
		LIVE1     & 0.58 (0.007)  & 3.57  & 39     & 7.14  (0.107)   & \textbf{5.01}  (11.82)   & 0.0218 (5.80)  & 2.87 (0.18) \\
		Set14     & 0.58 (0.013)  & 2.30  & 14     & 6.74  (0.605)   & 26.29 (147.77)  & 0.1421 (194.00)  & 4.71 (2.69) \\
		BSD100    & 0.60 (0.010)  & 1.54  & 100    & 6.94  (0.258)   & 20.01 (84.39)   & 0.0801 (97.13)  & 3.09 (0.76) \\
		DIV2K     & 0.53 (0.011)  & 28.35  & 100    & 7.02 (0.748)  & 23.64  (131.93)  & 0.0925 (47.01)  & 3.18 (2.05) \\
		\hline
		%		LIU-4K 25  & 0.47 (0.006)  & 82.94 & 200    & 7.27 (0.120)  & 26.33 (123.54) & 0.1132 (66.48) & 3.13 (0.52) \\
		LIU4K    & \textbf{0.92} (0.0298) &   \textbf{132.79} & \textbf{200} &  \textbf{7.43} (0.039) &  15.98 (73.86)  &  \textbf{0.0036} (32.02) &  \textbf{2.39} (0.24) \\
		\hline     
	\end{tabular}
	\vspace{-1mm}
\end{table*}

The summarization of all these datasets are provided in Table~\ref{tab:data_summary}.
\vspace{-3mm}

\begin{table*}[htbp]
	\caption{An Overview of Existing Works on Non-Deep JPEG Artifacts Removal.}
	\vspace{-2mm}
	\label{tab:nondeep}
	\scriptsize
	\centering
	\begin{tabular}{cccccm{3.5cm}}
		\hline
		Method &  Published & Category & Inference Model & Priors / Side Information & Basic Idea \\
		\hline
		\hline
		Minami and Zakhor & TCSVT-1995~\cite{texture_preservation} & Filter & \tabincell{l}{Linear model (constrained \\   quadratic programming)} & \tabincell{l}{Mean squared difference \\ of slope (MSDS)}   & Reducing the expected MSDS \\
		\hline
		\tabincell{c}{Deblocking Fliter} & TCSVT-2012~\cite{deblock} & Filter & \tabincell{c}{ Hand-crafted metrics \\
			based boundary classification} & \tabincell{c}{Coding information \\ (PU/TU, intra mode, \\ motion vector \textit{etc.})} & \tabincell{l}{Divide blocking boundaries into \\  different types and accordingly \\ choose diffident  kinds of deblocking \\ filters.
		} \\	
		%		\hline
		%		 Adaptive Loop Filtering & JSTSP-2013~\cite{ALF} & Filter & Wiener-based adaptive filter & Coding modes & \tabincell{l}{Deciding suitable filter coefficients \\ by the encoder} \\
		\hline
		Field of Expert & TIP-2007~\cite{SunC07} & Probabilistic prior & \tabincell{c}{MAP framework \\ High-order Markov model} & Quantization tables & \tabincell{l}{ The original image is modeled as a \\ high order MRF with learned \\ potential functions} \\
		\hline
		\tabincell{c}{Transform Domain \\ Non-Local Similarity} & \tabincell{c}{ICME-2012~\cite{zhang12} \\ TIP-2013~\cite{zhang13}} & Probabilistic prior & \tabincell{c}{MAP framework \\ Adaptive parameter selection} & Nonlocal similarity & \tabincell{l}{The decoded coefficient and its \\ nonlocal estimation are fused \\ adaptively} \\
		\hline
		\tabincell{c}{Low-Rank Minimization} & DCC-2013~\cite{dcc_low_rank} & Probabilistic prior & \tabincell{c}{Patch clustering \\
			Singular value thresholding} & \tabincell{c}{Local sparsity \\ Low-rank prior} & \tabincell{l}{Similar patches are clustered \\ and reconstructed by low-rank \\ minimization} \\
		\hline
		\tabincell{c}{Sparse Coding with \\ Total Variation} & TSP-2014~\cite{learned_dict} & Probabilistic prior & \tabincell{c}{Sparse coding} & \tabincell{c}{Sparsity prior\\ Total variations} & \tabincell{l}{Combination of sparse \\ representation and total variations} \\
		\hline
		\tabincell{c}{Dual Domain \\ Sparse Representation} & \tabincell{c}{CVPR-2015~\cite{dual_domain_external} \\ TIP-2016~\cite{dual_domain_external2}} & Probabilistic prior & \tabincell{l}{Sparse coding} & \tabincell{c}{Spatial domain\\ DCT domain \\ External data} & \tabincell{l}{Sparse representations \\ jointly in dual domains \\ augmented by external data } \\
		\hline
		\tabincell{l}{SA-DCT Transform} & TIP-2007~\cite{sa_dct} & Probabilistic prior & \tabincell{l}{Wiener filtering \\ in SA-DCT domain} & \tabincell{l}{SA-DCT transform\\ Structural constraint} & \tabincell{l}{The transform uses \\ adaptive supports which \\ leads to better edge reconstruction} \\
		\hline
		\tabincell{c}{Adaptive Low-Rank \\ Minimization} & TIP-2016~\cite{adaptive_low_rank} & Probabilistic prior & \tabincell{c}{Patch clustering \\
			Singular value thresholding} & \tabincell{c}{Local sparsity \\ Low-rank prior \\ Transform coefficient variance \\ Quantization step} & \tabincell{l}{The thresholds in SVT are \\ 
			adaptively determined} \\
		\hline
	\end{tabular}
	\vspace{-4mm}
\end{table*}

\begin{table*}[htbp]
	\caption{An Overview of Existing Works on Deep Compression Artifacts Removal.}
	\vspace{-2mm}
	\label{tab:deep}
	\scriptsize
	\centering
	\begin{tabular}{ccccm{6.2cm}}
		\hline
		Method &  Published & Inference Model & Priors / Side Information & Basic Idea \\
		\hline
		\hline
		\tabincell{c}{Artifacts Reduction \\ CNN} & ICCV-2015~\cite{ARCNN}  & \tabincell{l}{Three-layer CNN} & \tabincell{l}{/} & \tabincell{l}{The first work introducing deep models to the topic} \\
		\hline
		\tabincell{c}{Trainable Nonlinear \\ Reaction Diffusion} & TPAMI-2017~\cite{TNRD_pami}  & \tabincell{c}{Trainable nonlinear \\ diffusion model} & \tabincell{c}{/} & \tabincell{l}{The proposed nonlinear diffusion  model is unrolling \\ into a deep network} \\
		\hline
		\tabincell{c}{D$^3$ Model} & CVPR-2016~\cite{d3}  & \tabincell{l}{Learned iterative shrinkage \\ and thresholding with DCT \\ layers} & \tabincell{c}{DCT domain constraint \\ Sparsity constraint} & \tabincell{l}{The proposed nonlinear diffusion model is unrolling into a \\ deep network} \\
		\hline
		\tabincell{c}{Denoising CNN} & TIP-2017~\cite{DnCNN}  & \tabincell{l}{CNN with residual learning \\ and batch normalization} & / & \tabincell{l}{The combination of residual learning, batch normalization, \\ and Adam optimization} \\
		\hline
		\tabincell{c}{Dual-Domain CNN} & ECCV-2016~\cite{DDCN}  & \tabincell{l}{A two-branch CNN} & Range of DCT
		coefficients & \tabincell{l}{A two-branch CNN works in pixel and DCT  domains and \\ finally aggregates their information} \\
		\hline
		\tabincell{c}{Residual Encoder-\\ Decoder Network} & NIPS-2016~\cite{rednet}  & \tabincell{l}{Encoder-decoder with \\ skip connections} & / & \tabincell{l}{An encoder-decoder with symmetric skip connections} \\
		\hline
		\tabincell{c}{Compression Artifact \\ Suppression CNN} & IJCNN-2017~\cite{cascnn}  & \tabincell{l}{Encoder-decoder with \\ skip connections} & \tabincell{l}{Multi-scale losses} & \tabincell{l}{An encoder-decoder constrained by multi-scale losses} \\
		\hline
		\tabincell{c}{One-to-Many \\ Network} & CVPR-2017~\cite{one_to_many}  & \tabincell{c}{ResNet \\ Shift-and-average strategy} & \tabincell{c}{Perceptual loss \\ Adversarial loss \\ JPEG loss} & \tabincell{l}{A ResNet takes input as random noise and a compressed \\ image, and  its output is constrained by three  losses} \\
		\hline
		\tabincell{c}{MemNet} & ICCV-2017~\cite{memnet}  & \tabincell{c}{DenseNet architecture \\ Memory block } & \tabincell{c}{Multi-supervision \\ Long-term memory} & \tabincell{l}{
			The network is stacked by memory blocks, consisting of a \\ recursive unit and a gate unit to learn explicit persistent \\ memories
		} \\
		\hline
		\tabincell{c}{DMCNN} & ICIP-2018~\cite{DMCNN}  & \tabincell{l}{A two-branch auto-encoder \\ with dilated convolution} & \tabincell{c}{DCT domain constraint \\ Multi-scale loss} & \tabincell{l}{
			It integrates the dual domain architecture (DCT and spatial \\ domains), DCT loss and multi-scale loss
		} \\
		\hline
		\tabincell{c}{Multi-level \\ Wavelet-CNN} & CVPRW-2018~\cite{Liu_2018_CVPR_Workshops}  & \tabincell{l}{Encoder-decoder with \\ Wavelet transforms} & \tabincell{c}{Wavelet signal structure} & \tabincell{l}{Wavelet transforms are introduced into CNN architecture} \\
		\hline
		\tabincell{c}{Dual Pixel-Wavelet \\ Domain Deep CNN} & CVPRW-2018~\cite{DPW-SDNet}  & \tabincell{c}{A two-branch CNN \\ with Wavelet transforms} & \tabincell{c}{Dual domains \\ Wavelet signal  structure} & \tabincell{l}{A two-branch CNN is constructed to make use of both \\ redudenancy in pixel and frequency domains} \\
		\hline
		\hline
		\tabincell{c}{VRCNN} & MMM-2017~\cite{VRCNN}  & \tabincell{c}{Variable-filter-size \\  Residue-learning} & \tabincell{c}{/} & \tabincell{l}{The designed CNN owns variable filter size to learn  the \\ residual between input and target frames} \\
		\hline
		\tabincell{c}{Deep CNN-based \\ Auto Decoder} & DCC-2017~\cite{DCAD}  & \tabincell{c}{ResNet} & \tabincell{c}{TU size} & \tabincell{l}{A ResNet is used for quality enhancement in the decoder end
		} \\
		\hline
		\tabincell{c}{Partition Mask CNN} & ICIP-2018~\cite{partition_mask}  & \tabincell{c}{ResNet} & \tabincell{c}{CU size} & \tabincell{l}{
			The CU size is utilized  and integrated with distorted decoded \\ frame} \\
		\hline
		\tabincell{c}{Residual High-way \\ CNN} & TIP-2018~\cite{highway_network}  & \tabincell{c}{Highway network} & \tabincell{c}{QP range} & \tabincell{l}{
			Residual highway CNNs trained 
			delicately for each QP range} \\
		\hline
		\tabincell{c}{MLSDRN} & DCC-2018~\cite{MLSDRN}  & \tabincell{l}{Multi-channel long-short \\ term dependency residual \\ network} & \tabincell{c}{Block boundary \\ Multi-channel} & \tabincell{l}{
			MLSDRN uses an update cell
			to adaptively store and select \\ the long-term and short-term dependency} \\
		\hline
		\tabincell{c}{Adversarial \\ Intra Coding} & ICASSP-2018~\cite{adversarial}  & \tabincell{c}{Multi-scale structure} & \tabincell{c}{Adversarial learning} & \tabincell{l}{
			A multi-level progressive refinement network with adversarial \\ learning} \\
		\hline
		\tabincell{c}{Decoder-Side \\ Scalable CNN} & ICME-2017~\cite{DSCNN}  & \tabincell{c}{Two-branch scalable CNN} & \tabincell{c}{/} & \tabincell{l}{
			The network has two branches. A group of switches will \\ control whether the  complicated one is activated.
		} \\
		\hline
		\tabincell{c}{Practical CNN} & ICIP-2018~\cite{PraticalCNN}  & \tabincell{c}{Compressed fixed \\ point CNN} & \tabincell{c}{QP} & \tabincell{l}{
			The network also takes QP as input. After training, the model \\ is  compressed and concerted into fixed  point format.
		} 
		\\
		\hline
		\tabincell{c}{Multi-Scale \\ Deep Decoder} & DCC-2018~\cite{MSDD}  & \tabincell{c}{CNN \\ Multi-scale LSTM} & \tabincell{c}{/} & \tabincell{l}{Each frame is fed into a CNN, then a multi-scale LSTM is \\ connected to fuse multi-frame redundancies.} 
		\\
		\hline
		\tabincell{c}{MF Quality Enhancer} & CVPR-2018~\cite{MFQE}  & \tabincell{c}{SVM classier \\ CNN-based alignment \\ CNN-based enhancer} & \tabincell{c}{Neighboring peak \\ quality frames} & \tabincell{l}{The neighboring high quality frames are fed into a CNN to \\ facilitate inference of enhanced frames.} 
		\\
		\hline
		\tabincell{c}{Separable CNN filter} & JVET-K0158~\cite{Jvet1}  & \tabincell{c}{SE block \\ Separable convolution} & \tabincell{c}{Normalized Y/U/V \\ Normalized QP} & \tabincell{l}{The network takes as input Normalized Y/U/V and QP and \\ consists of SE blocks and separable convolutions.} 
		\\
		\hline
		\tabincell{c}{Dense Residual Network} & VCIP-2018~\cite{dense_residual}  & \tabincell{c}{Dense shortcuts \\ Residual learning \\ Bottleneck layer} & \tabincell{c}{/} & \tabincell{l}{
			The network consists of dense 
			shortcuts, residual learning, \\ 
			and bottleneck layers.
		}
		\\
		\hline
		\tabincell{c}{CU Classification} & VCIP-2018~\cite{CUC}  & \tabincell{c}{Multiple variable-filter-size \\  residue-learning} & \tabincell{c}{CU classification } & \tabincell{l}{
			A classifier is employed to decide whether to use VRCNN-ext \\ for each coding unit.
		}
		\\
		\hline
		\tabincell{c}{Progressive Rethinking \\ Network} & ICIP-2019~\cite{PRN}  & \tabincell{c}{Progressive Rethinking \\ Block and Network } & \tabincell{c}{Multi-scale mean \\ value of CUs} & \tabincell{l}{
			The progressive rethinking network is built to take multi-scale \\ mean value of coding
			units as side information.
		}
		\\
		\hline
		\tabincell{c}{Coding Prior based \\ High Efficiency
			Restoration} & ICIP-2019~\cite{CPHER}  & \tabincell{c}{Weight Normalization} & \tabincell{c}{Unfiltered frame \\ Prediction frame} & \tabincell{l}{
			An EDSR-like network takes the unfiltered and prediction \\ frames as side information and is trained with weight \\ normalization.
		}
		\\
		\hline
		\tabincell{c}{Content-Aware CNN} & TIP-2019~\cite{CACNN}  & \tabincell{c}{Context-based \\ model selection} & \tabincell{c}{Clusters based on \\ quality ranking} & \tabincell{l}{
			The discriminative model is learned to analyze the region \\ 
			content for model selection. An iterative training is proposed \\
			to label filter categories and fine-tune CNN models.
		}
		\\
		\hline
	\end{tabular}
	\vspace{-4mm}
\end{table*}

\subsection{LIU4K Dataset}

The main characteristics of the LIU4K dataset and previous datasets serving for image restoration are listed in Table~\ref{tab:quality_metrics}.
LIU4K has several unprecedented superiorities as follows,
\begin{itemize}
	\item \textbf{High-resolution definition}. Compared to previous datasets, the resolution of the images in our dataset is 2848$\times$4288, larger than those in previous datasets, which offers abundant materials for testing and evaluating the performance in 4K/8K display devices. 
	\item \textbf{Large-scale}. Our dataset is large-scale. Our training and testing images include 1,541 and 200 4K images, much more than those in previous datasets. Thus, training and evaluation based on LIU4K are more comprehensive and balanced.
	\item \textbf{Diversified and complex signals}.
	As shown in Table~\ref{tab:quality_metrics}, our dataset achieves the best results in entropy-driven non-reference metrics, which demonstrates its signal diversity and complexity.
	\item \textbf{High visual quality}. LIU4K wins in general purpose non-reference metrics (except for Kodak and LIVE1, which are the training set of some metrics) as shown in Table~\ref{tab:quality_metrics}, which confirms its high visual quality.
\end{itemize}

We perform statistical comparisons to demonstrate the superiority of LIU4K dataset.
Entropy, BPP (Bits Per Pixel) and PPI (Pixels Per Image) are used to indicate the amount of information included in each dataset. Three \wh{non-reference} image quality assessment metrics are utilized to assess the perceptual image quality, including Entropy, natural image quality evaluator (NIQE)~\cite{NIQE}, blind/referenceless image spatial quality evaluator (BRISQE)~\cite{BRISQE}, and entropy-based image quality assessment (ENIQA).
\wht{The entropy is estimated following the most primitive calculation based on per-pixel independent distribution~\cite{entropy}.}
\wht{The bits used to calculate BPP values are estimated by the compressing the gray version of an image into a PNG image.}
The work in~\cite{MA20171} has shown that, the non-reference image quality assessment metrics are highly correlated to human perception and are superior to some full-referenced measures in visual quality. In our work, we calculate values of NIQE, BRISQE, and ENIQA with the codes provided by their authors with the default settings. For NIQE, BRISQE and ENIQA, small values indicate better image qualities.

As listed in Table~\ref{tab:quality_metrics}, LIU4K is more large-scale than previous datasets in scale and resolution. From the perspective of information theory, the images in LIU4K are more informative. Its mean BPP and entropy values are greater, which means that the dataset contains more information. 
For perceptual image quality assessment, LIU4K also achieves very competitive scores in BRISQUE, ENIQA, and NIQE.
Note that, the values of BRISQUE and ENIQA of LIU4K are worse than those of Kodak and LIVE1, since the metrics are trained on these two datasets and their scales are relative small.
These assessments indicate that images in LIU4K are of relatively high perceptual quality and suitable for image restoration tasks.
\vspace{-2mm}

\section{Algorithm Survey}

The approaches designed for compression artifacts removal, namely loop filters in codecs, are proposed in literatures. 
There are four categories in our review: filter-based methods, probabilistic-prior based methods, deep learning-based JPEG artifacts removal methods, and deep learning-based loop filter \wh{methods}.
\wht{The first and last two categories are summarized in Table~\ref{tab:nondeep} and~\ref{tab:deep}, respectively. 
We will review the four categories and then briefly summarize their technical improvements. 
%After that, a special attention is paid to the network architecture improvement.
Note that, the technologies discussed in our work can be applied without the change of the existing pipeline of codecs.}
\vspace{-3mm}

\subsection{Filtering-based method}
The earliest methods~\cite{start_work,Ramamurthi1986NonlinearSP} perform filtering operations to remove compression artifacts. Later approaches~\cite{texture_preservation,HEVC} attempt to infer the parameters of filtering operations adaptively. 
Minami and Zakhor~\cite{texture_preservation} observed that quantization of the DCT coefficients of two neighboring blocks increases the expected value of the mean squared difference of slope (MSDS) between the slope across two adjacent blocks, and the average value between the boundary slopes of each of the two blocks. Thus, a constrained quadratic programming problem is built to reduce the expected value of this MSDS to extent to decrease the blocking effect while preserving texture details. \wht{In HEVC, an in-loop deblocking filter is specially designed~\cite{deblock} to reduce the blocking artifacts between coding units. The picture is divided into $8\times8$ blocks and boundaries on the $8\times8$ grid are classified by a series of metrics. Different levels of deblocking operations are later performed on the boundaries according to their types.}

\vspace{-3mm}

%\noindent \textbf{Minami and Zakhor}. Minami and Zakhor~\cite{texture_preservation} observed that quantization of the DCT coefficients of two neighboring blocks increases the expected value of the mean squared difference of slope (MSDS) between the slope across two adjacent blocks, and the average value between the boundary slopes of each of the two blocks. Thus, a constrained quadratic programming problem is built to reduce the expected value of this MSDS to extent to decrease the blocking effect while preserving texture details.
%
%\noindent \textbf{Adaptive Loop Filtering}. In the later video coding standards, \textit{e.g.}, HEVC~\cite{HEVC}, the strength of deblocking filter is adaptively decided. 
%Adaptive loop filtering (ALF)~\cite{ALF} in HEVC minimizes the mean square error between original samples and decoded ones with Wiener-based adaptive filter that is derived by the encoder.
%To achieve better coding efficiency, for luma signals, different filters are applied to regions or blocks in a picture based on the coding modes.

%\subsection{Probabilistic-Prior Methods}

%Besides the filter-based methods, there are approaches based on probability estimation of image prior models.

\subsection{Probabilistic-prior methods.} 
\wht{Some successive approaches are based on probability estimation of image prior models.
Based on their basic models, these methods can be further categorized into Markov random field~\cite{SunC07}, non-local similarity~\cite{zhang12,zhang13}, low-rank minimization~\cite{dcc_low_rank,zhang13}, sparse coding~\cite{learned_dict,dual_domain_external, dual_domain_external2}, and adaptive DCT transformation~\cite{sa_dct}.
In~\cite{SunC07}, the distortion term is modeled as additive, spatially correlated Gaussian noise, and the original image is depicted as a high order Markov random field based on the fields of experts framework.
Non-local based methods~\cite{zhang12,zhang13} consider similar blocks to be potentially correlated, estimate the overlapped-block transform coefficients, and remove compression noise from non-local similar blocks. For low-rank based methods, Ren \textit{et al.}~\cite{dcc_low_rank} performed patch clustering and low-rank minimization simultaneously to make use of both local sparsity and non-local similarity. A successive work~\cite{zhang13} selects thresholds adaptively for each group of similar patches based on compression noise levels and decomposed singular values. In~\cite{sa_dct}, a new shape adaptive DCT transform is proposed for image compression artifacts reduction. 
}

\begin{figure*}[t]
	\centering
	\subfigure{
		\includegraphics[width=18.1cm]{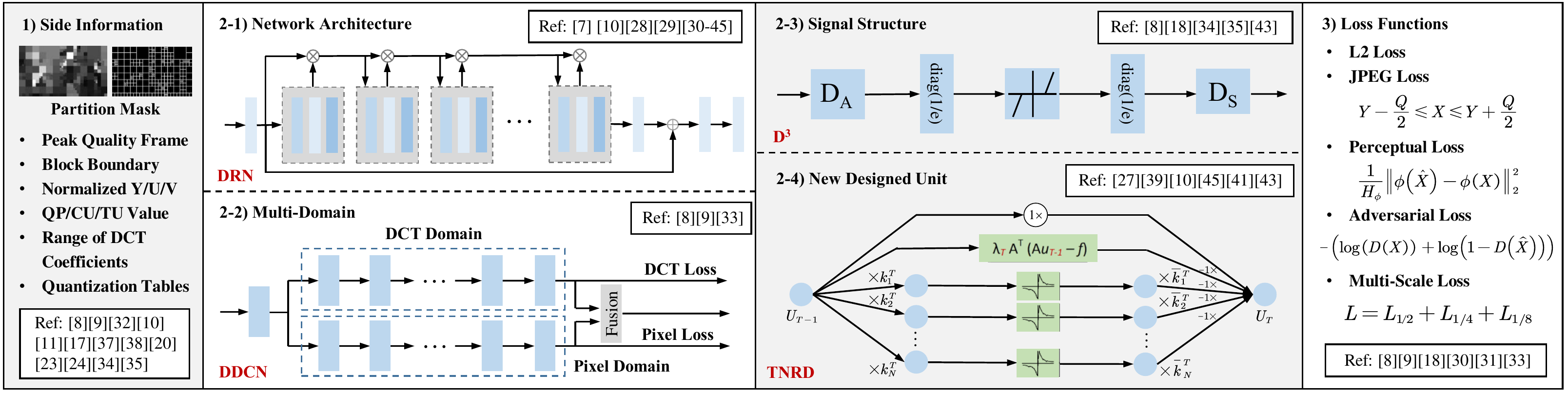}}
	\vspace{-2mm}
	\caption{The technical improvement route for deep learning-based compression artifacts removal and loop filter of codecs.}
	\vspace{-5mm}
	\label{fig:improvement}
\end{figure*}

\begin{figure}[!h]
	\centering
	\subfigure{
		\includegraphics[width=7.5cm]{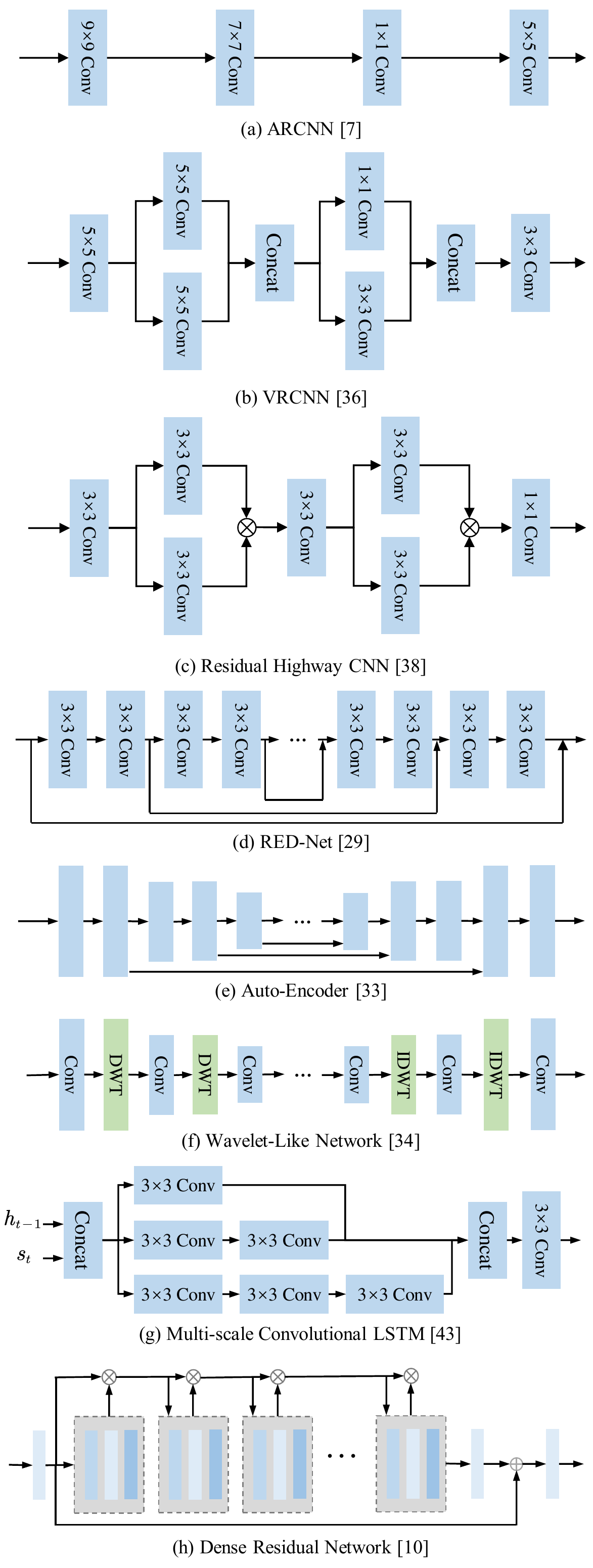}}
	\vspace{-2mm}	
	\caption{The network improvement route for compression artifacts removal and loop filter of codecs, \whtt{where the multiplication sign in the circle in~(c) denotes the element-wise multiplication operation.}
	}
	\vspace{-7mm}
	\label{fig:network}
\end{figure}

\subsection{Deep learning-based JPEG artifacts removal}
\begin{figure}[t]
	\centering
	\subfigure{
		\includegraphics[width=8cm]{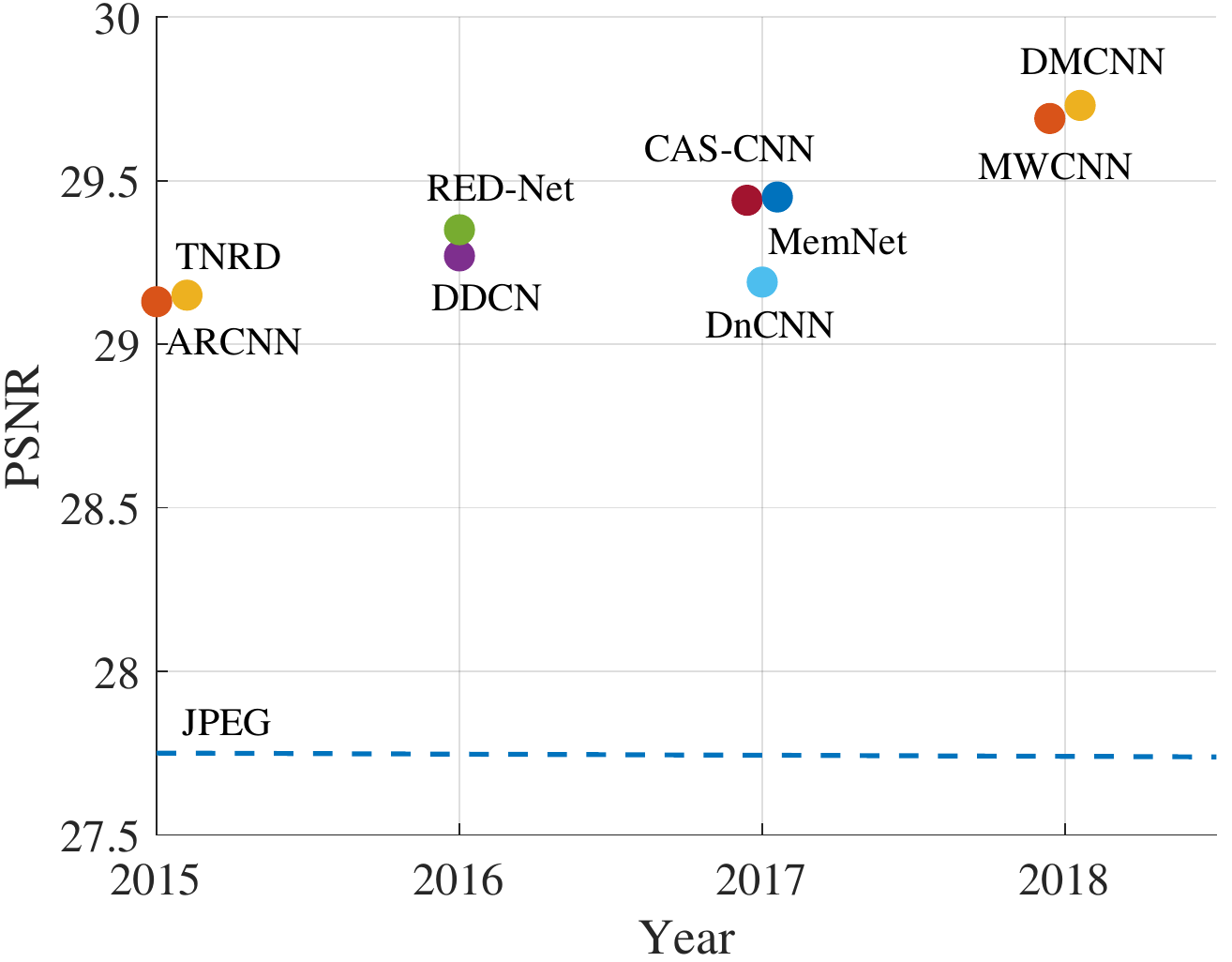}}
	\caption{ Recent evolution of deep learning-based JPEG artifacts removal. We can observe significant performance (PSNR) improvement since deep learning entered the scene in 2015. The performances showed here are directly quoted from the published papers.
	}
	\vspace{-6mm}
	
	\label{fig:jpeg_performance}
\end{figure}

\wht{Deep learning-based methods largely improve the restoration capacity of the data-driven methods. 
ARCNN~\cite{ARCNN} is the seminal work and takes the architecture of a three-layer CNN.
Deep Dual-Domain (D$^3$)~\cite{d3} is the first work to introduce the DCT-domain priors to facilitate JPEG artifacts removal. It combines both the strong learning capacity of deep networks, as well as the problem-specific knowledge of JPEG artifacts removal.

Successive works proceed into two main streams: better network architectures~\cite{DnCNN,cascnn,memnet} and better utilization of DCT domain information~\cite{DMCNN,DDCN}. 
Many advanced networks are constructed to model the rich dependencies of deep features. 
Residual Encoder-Decoder Network (RED-Net)~\cite{rednet} and Compression Artifact Suppression CNN (CAS-CNN)~\cite{cascnn} utilize  deep encoding-decoding frameworks with symmetric convolutional-deconvolutional layers. Tai \textit{et al.}~\cite{memnet} constructed a deep persistent memory network. The memory blocks consist of a recursive unit and a gate unit to keep memories. The former extracts multi-level representations from the last input feature while the latter learns to control the ratio between the memory and current input.
Dual-Domain Multi-Scale CNN (DMCNN)~\cite{DMCNN} integrates the dual domain and  auto-encoder style networks with dilated convolutions to have very large receptive fields and eliminate the banding effects. 
In~\cite{Liu_2018_CVPR_Workshops}, wavelet transforms are introduced into CNN architecture for a better trade-off between the receptive field size and computational efficiency.
In~\cite{DPW-SDNet}, a two-branch CNN is constructed to handle the restoration in the pixel and discrete wavelet domains.

Besides network improvement, some works try to embed traditional priors or constraints into deep networks, \textit{e.g.} sparsity~\cite{d3}, nonlinear diffusion~\cite{TNRD_pami}, multi-scale constraint~\cite{cascnn,DMCNN}, and wavelet signal structures~\cite{Liu_2018_CVPR_Workshops,DPW-SDNet}. In one-to-many network~\cite{one_to_many}, adversarial learning is introduced to facilitate generating visually pleasing restoration results.
A summary of the performance comparison of typical deep learning-based JPEG artifacts removal reported by their works is presented in Fig.~\ref{fig:jpeg_performance}.
}

\subsection{Deep Learning-Based Loop Filter} 

Besides JPEG, deep learning techniques are also applied to latest codecs, \textit{e.g.} HVEC, as a post-processor. 
\wht{Beyond the improvement directions embodied in JPEG artifacts removal, deep-learning based loop filter considers more on handling the degradation caused by variable-size partition and utilizing the side information from codecs.}
\wht{Variable-Filter-Size Residue-Learning CNN (VRCNN)~\cite{VRCNN} is the pioneering work.
The designed CNN owns variable filter size to learn the residual between input and target frames.
Successive works also proceed into two classes: better network and better side information.
Zhang \textit{et al.}~\cite{highway_network} proposed a residual highway convolutional neural network (RHCNN) for in-loop filter of HEVC. 
In~\cite{MSDD}, Wang \textit{et al.} proposed a multi-scale LSTM to fuse multi-frame redundancies along temporal dimension to acquire the fused feature.
Meng \textit{et al.}~\cite{MLSDRN} proposed a multi-channel long-short term dependency residual network to simulate the mechanism of human memory update and introduced an update cell which learns to store and select the long-term and short-term dependency adaptively.
Li \textit{et al.}~\cite{CNNClassifyPost} presented a dynamic classification mechanism. An up-to-one byte flag indicates complexity of video content and quality of each frame.
In~\cite{DSCNN}, Yang \textit{et al.} designed a scalable deep CNN to reduce distortion of both I and B/P frames in HEVC.  is proposed. It has two branches and a group of switches to control whether DS-CNN-B branch is activated based on the resource state. In~\cite{PraticalCNN}, Song \textit{et al.} developed a CNN that can enhance compressed videos of different qualities with low redundancy. In~\cite{MFQE}, Yang \textit{et al.} explored to enhance the compressed video frames using the neighboring high quality frames.
A novel multi-frame convolutional neural network is built for compressed video enhancement.
In~\cite{Jvet1}, Hashimoto  \textit{et al.} proposed a CNN with squeeze and excitation block and spatial separable convolution for deblocking. 
In~\cite{dense_residual}, Wang \textit{et al.} proposed a dense residual convolutional neural network (DRN). 
In this network, dense shortcuts and residual learning are combined.
Bottleneck layers are injected into each DRN to save the computational resources while adaptively fusing the hierarchical features.

Various kinds of side information is designed for more effective post-processing of \wht{compression artifacts reduction}.
This side information includes: the compression parameters from coding tree units (CTU)~\cite{MMSCNN}, partition mask of CTU~\cite{partition_mask}, QP parameter~\cite{highway_network}, block boundary~\cite{MLSDRN}, complexity~\cite{CNNClassifyPost}, peak quality frames and optical flow~\cite{MFQE}, and normalized Y/U/V and normalized QP~\cite{Jvet1}, \textit{etc}.
}

\wht{\subsection{Technical Improvement Summary}
The typical improvement route of deep learning-based compression \wht{artifacts reduction} is summarized in Fig.~\ref{fig:improvement}. 
Three aspects of improvements are included: 
\textit{side information utilization}, \textit{e.g.} injecting the partition mask of CTU~\cite{partition_mask} as input; 
\textit{network improvement}, \textit{e.g.} dense residual network~\cite{dense_residual};  
and \textit{novel loss function}, \textit{e.g.} adversarial loss~\cite{one_to_many}.
For the network improvement, all methods are improved in four directions:
1) \textit{network architecture improvement} (summarized more specifically in Fig.~\ref{fig:network}); 
2) \textit{multi-domain network}, \textit{e.g.} DMCNN~\cite{DMCNN}; 
3) \textit{signal structure embedding}, \textit{e.g.} D3~\cite{d3}; 
4) \textit{new unit design}, \textit{e.g.} TNRD~\cite{TNRD_pami}.
In the next section, we will benchmark these methods using the unified protocols.
}

\section{Algorithm Benchmarking}
\label{sec:benckmark}

With the rich resources provided by LIU4K, 
we evaluate 9 representative state-of-the-art algorithms: 
Shape-Adaptive DCT (SA-DCT)~\cite{sa_dct}, 
%Regression Tree Fields (RTF)~\cite{RTF},
Artifacts Removal CNN (ARCNN)~\cite{ARCNN},
Trainable Nonlinear Reaction Diffusion (TNRD)~\cite{TNRD_pami},
Denoising CNN (DnCNN)~\cite{DnCNN},
%Deep Dual-Domain restoration (D3)~\cite{d3},
%BlockCNN~\cite{Maleki_2018_CVPR_Workshops},
Persistent Memory Network~MemNet~\cite{memnet},
Dual-domain Multi-scale CNN (DMCNN)~\cite{DMCNN},
Multi-Level Wavelet-CNN (MWCNN)~\cite{Liu_2018_CVPR_Workshops},
Variable-filter-size Residue-learning CNN (VRCNN)~\cite{VRCNN},
and Progressive Rethinking Network (PRN)~\cite{PRN}.
Our selected baselines try to cover most of the representative methods.
The first one is a traditional non-deep method.
The successive six methods are deep learning-based \wht{JPEG artifacts reduction} methods. 
The last two are deep learning-based loop filter methods. 
We apply most of learning-based methods to restorations of images compressed by JPEG and HEVC.
For \wht{JPEG artifacts reduction}, we train the models on the training of LIU4K. 
For loop filters, the models are trained on the training sets of both BSD500 and LIU4K.
\wht{Note that, the source codes of SA-DCT and TNRD provided by the authors only support removing \wht{JPEG artifacts} with quality factors 10, 20, 30, 40 and 10, 20, 30, respectively. Thus, for these two methods, we only compare their performances in these cases.}
We also add the residual learning in our implemented ARCNN for fast training and comparison.

\subsection{Advanced Training Strategies}

In our benchmarking, we also make efforts in generalizing some constraints and methods of \wht{JPEG artifacts reduction} to the general compression artifacts \wht{reduction}.
\\
\noindent \textbf{Dense DCT Domain Constraint}. 
For some codecs, \textit{e.g.} HEVC, the partitioned block sizes are not the same all the time. Thus, the original DCT branch constraint that regularizes reconstruction of fixed-sized blocks in JPEG artifacts removal cannot be utilized. Thus, we slightly change our DCT branch design:
\begin{itemize}
	\item Location-independence. 
	We do not pose the constraint based on block partitions. Instead, we impose the DCT constraint densely for every pixel location.
	\item Variable-size DCT constraint. To handle variable block sizes in HEVC codecs, we extend the DCT branch into multiple branches. Each branch is responsible for constructing DCT constraints at a certain scale.
\end{itemize}
In summary, our DCT constraint takes several variable-block-size DCT transformation paths densely in every location. 
\\
\noindent \textbf{Gradual Expanding Patch Size}.
The DCT branch is not stable during training. To make our training more effective, we first use small patches to train our network and then enlarge the patch size gradually. Let $P$ and $e$ denote the patch size and epoch. The size of training patches during train a restoration model for JPEG is set as follows,
\begin{equation}  
P=\left\{  
\begin{array}{rc}  
56, & e \in \left[1, 6\right],    \\  
112, & e \in \left[7, 9\right],   \\  
168, & e \in \left[10, 12\right], \\  
224, & e \in \left[13, 15\right], \\
256, & e \in \left[16, 60\right]. \\  
\end{array}  
\right.  
\end{equation}
For HEVC post-processing, all interval bounds for $e$ are multiplied by 5.
This strategy leads to the better constraint of the DCT branch and also offers a better performance.

\noindent \textbf{Learning with Mixed Batches}.
For the methods with high complexities, 
it is impossible to train the models with both a large patch size and a large batch size at the same time.
Thus, we propose to apply training with mixed batches, the combination of (large patch, small batch) and (small patch, large batch).
Therefore, with the limited GPU memory resources, the network training can be stabilized by the large batch size, and at the same time, the model can also learn the information from a large context with the large patch size. 
In our benchmark, we train MemNet and PRN with this strategy.
\vspace{-4mm}

\newcommand{\gwidth}{3.3cm}

\begin{figure*}[htbp]
	\flushright
	\subfigure{
		\includegraphics[width=\gwidth]{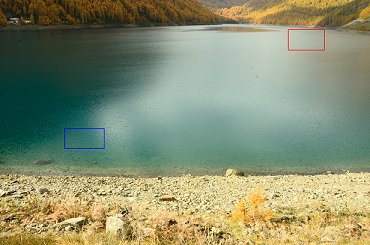}
	}
	\hspace{-3mm}	
	\subfigure{
		\includegraphics[width=\gwidth]{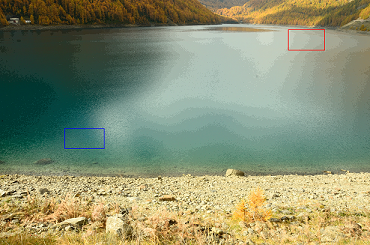}
	}
	\hspace{-3mm}
	\subfigure{
		\includegraphics[width=\gwidth]{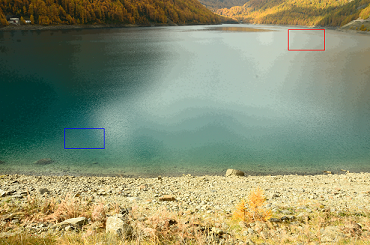}
	}
	\hspace{-3mm}
	\subfigure{
		\includegraphics[width=\gwidth]{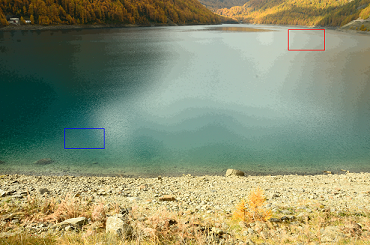}
	}
	\hspace{-3mm}
	\subfigure{
		\includegraphics[width=\gwidth]{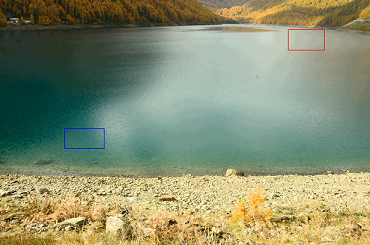}
	}
	\\
	\vspace{-2mm}
	\subfigure{
		\includegraphics[width=\gwidth]{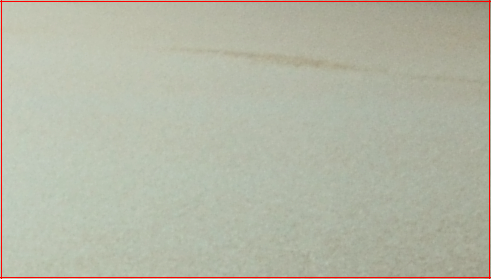}
	}
	\hspace{-3mm}
	\subfigure{
		\includegraphics[width=\gwidth]{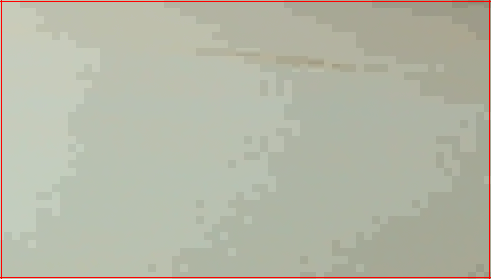}
	}
	\hspace{-3mm}
	\subfigure{
		\includegraphics[width=\gwidth]{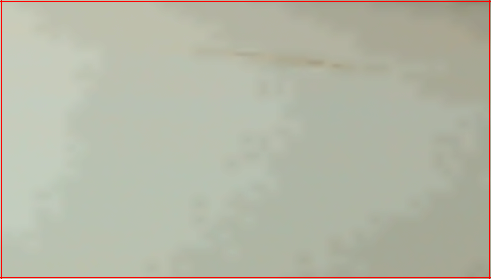}
	}
	\hspace{-3mm}
	\subfigure{
		\includegraphics[width=\gwidth]{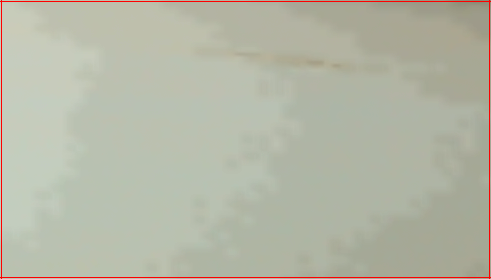}
	}
	\hspace{-3mm}
	\subfigure{
		\includegraphics[width=\gwidth]{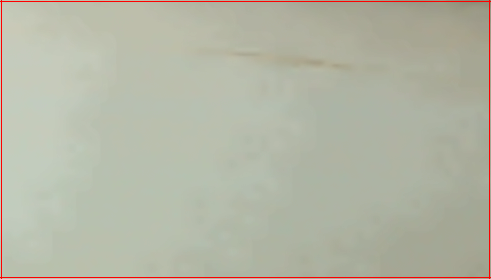}
	}
	\\
	\vspace{-2mm}
	\setcounter{subfigure}{0}
	\subfigure[Original]{
		\includegraphics[width=\gwidth]{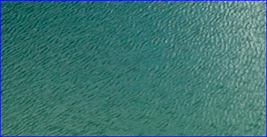}
	}
	\hspace{-3mm}
	\subfigure[JPEG]{
		\includegraphics[width=\gwidth]{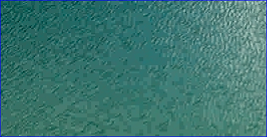}
	}
	\hspace{-3mm}
	\subfigure[ARCNN~\cite{ARCNN}]{
		\includegraphics[width=\gwidth]{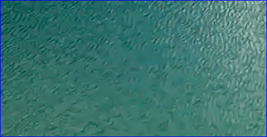}
	}
	\hspace{-3mm}
	\subfigure[VRCNN~\cite{VRCNN}]{
		\includegraphics[width=\gwidth]{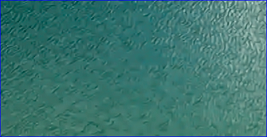}
	}
	\hspace{-3mm}
	\subfigure[DnCNN~\cite{DnCNN}]{
		\includegraphics[width=\gwidth]{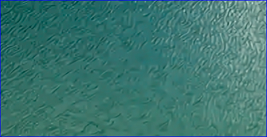}
	}
	\\	
	\subfigure{
		\includegraphics[width=\gwidth]{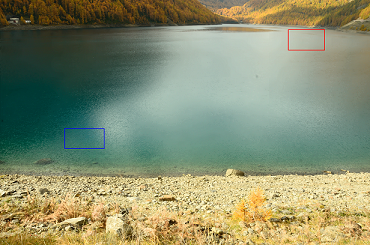}
	}
	\hspace{-3mm}
	\subfigure{
		\includegraphics[width=\gwidth]{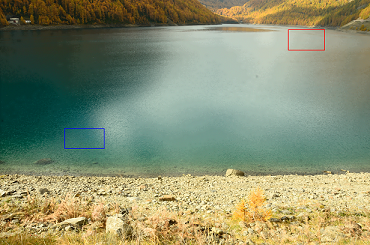}
	}
	\hspace{-3mm}
	\subfigure{
		\includegraphics[width=\gwidth]{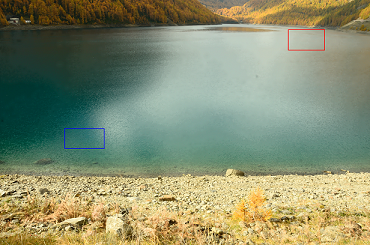}
	}
	\hspace{-3mm}
	\subfigure{
		\includegraphics[width=\gwidth]{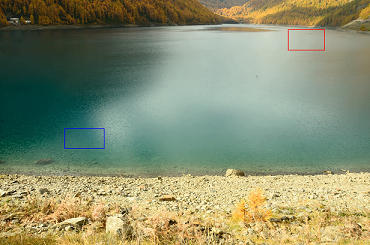}
	}
	\\
	\vspace{-2mm}
	\subfigure{
		\includegraphics[width=\gwidth]{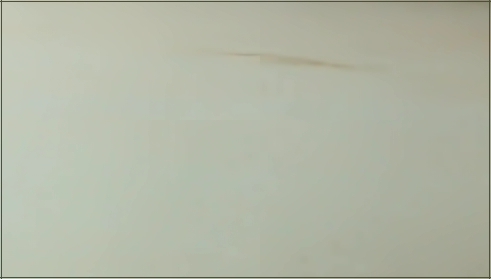}
	}
	\hspace{-3mm}
	\subfigure{
		\includegraphics[width=\gwidth]{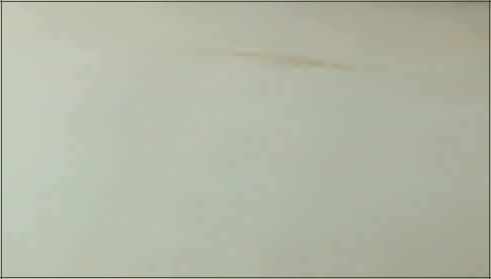}
	}
	\hspace{-3mm}
	\subfigure{
		\includegraphics[width=\gwidth]{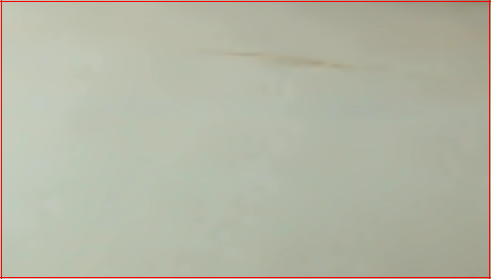}
	}
	\hspace{-3mm}
	\subfigure{
		\includegraphics[width=\gwidth]{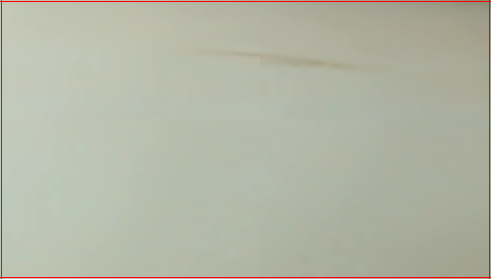}
	}
	\\
	\vspace{-2mm}
	\setcounter{subfigure}{5}
	\subfigure[MWCNN~\cite{Liu_2018_CVPR_Workshops}]{
		\includegraphics[width=\gwidth]{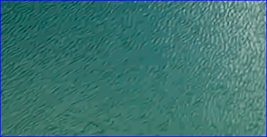}
	}
	\hspace{-3mm}
	\subfigure[MemNet~\cite{memnet}]{
		\includegraphics[width=\gwidth]{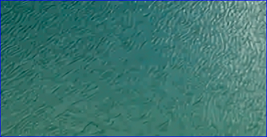}
	}
	\hspace{-3mm}
	\subfigure[PRN~\cite{PRN}]{
		\includegraphics[width=\gwidth]{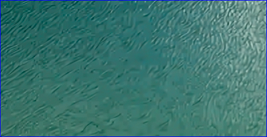}
	}
	\hspace{-3mm}
	\subfigure[DMCNN~\cite{DMCNN}]{
		\includegraphics[width=\gwidth]{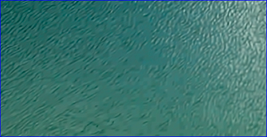}
	}
	\\
	\vspace{-1mm}
	\caption{Examples of restored results on a compressed image by HM from LIU4K (QF=10).}
	\vspace{-2mm}
	\label{fig:visual_comparison}
\end{figure*}

\begin{figure*}[htbp]
	\flushright
	\subfigure{
		\includegraphics[width=\gwidth]{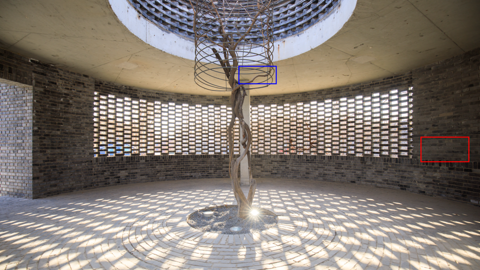}
	}
	\hspace{-3mm}	
	\subfigure{
		\includegraphics[width=\gwidth]{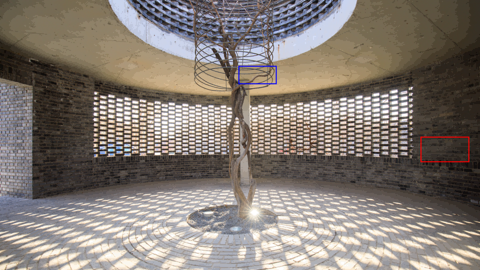}
	}
	\hspace{-3mm}
	\subfigure{
		\includegraphics[width=\gwidth]{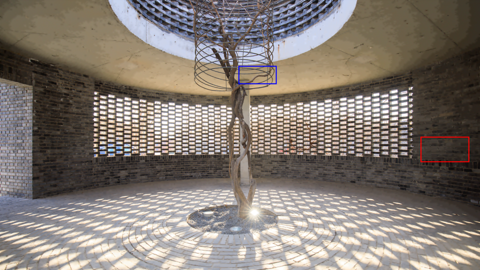}
	}
	\hspace{-3mm}
	\subfigure{
		\includegraphics[width=\gwidth]{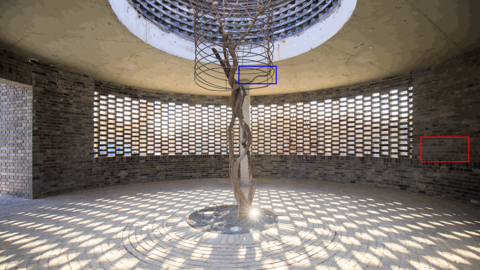}
	}
	\hspace{-3mm}
	\subfigure{
		\includegraphics[width=\gwidth]{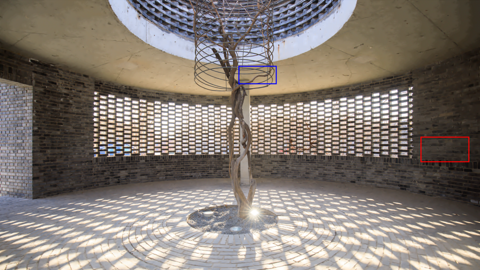}
	}
	\\
	\vspace{-2mm}
	\setcounter{subfigure}{0}
	\subfigure{
		\includegraphics[width=\gwidth]{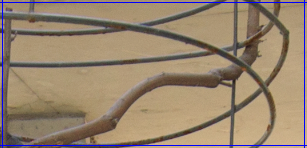}
	}
	\hspace{-3mm}
	\subfigure{
		\includegraphics[width=\gwidth]{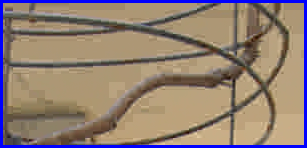}
	}
	\hspace{-3mm}
	\subfigure{
		\includegraphics[width=\gwidth]{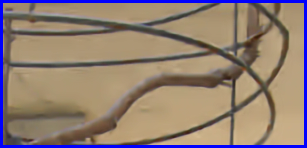}
	}
	\hspace{-3mm}
	\subfigure{
		\includegraphics[width=\gwidth]{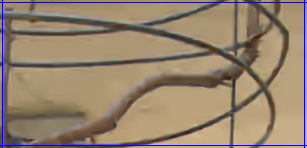}
	}
	\hspace{-3mm}
	\subfigure{
		\includegraphics[width=\gwidth]{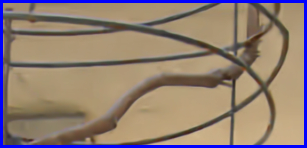}
	}
	\\
	\vspace{-2mm}
	\setcounter{subfigure}{0}
	\subfigure[Original]{
		\includegraphics[width=\gwidth]{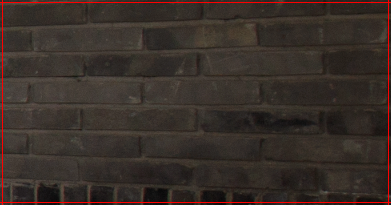}
	}
	\hspace{-3mm}
	\subfigure[JPEG]{
		\includegraphics[width=\gwidth]{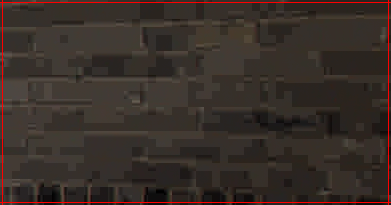}
	}
	\hspace{-3mm}
	\subfigure[ARCNN~\cite{ARCNN}]{
		\includegraphics[width=\gwidth]{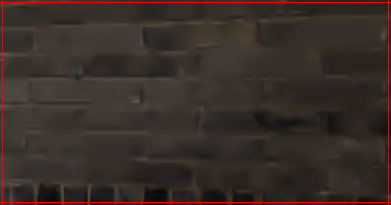}
	}
	\hspace{-3mm}
	\subfigure[VRCNN~\cite{VRCNN}]{
		\includegraphics[width=\gwidth]{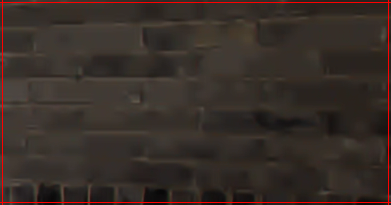}
	}
	\hspace{-3mm}
	\subfigure[DnCNN~\cite{DnCNN}]{
		\includegraphics[width=\gwidth]{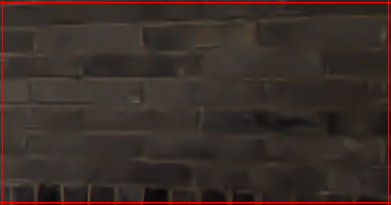}
	}
	\\
	\subfigure{
		\includegraphics[width=\gwidth]{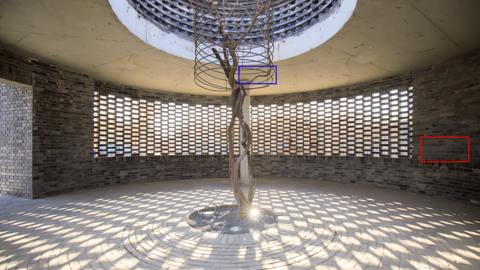}
	}
	\hspace{-3mm}
	\subfigure{
		\includegraphics[width=\gwidth]{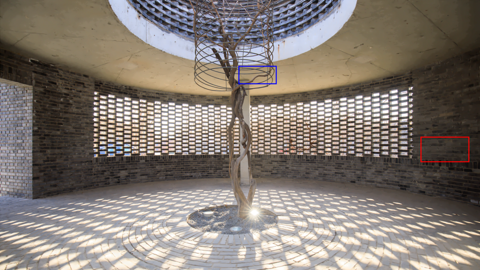}
	}
	\hspace{-3mm}
	\subfigure{
		\includegraphics[width=\gwidth]{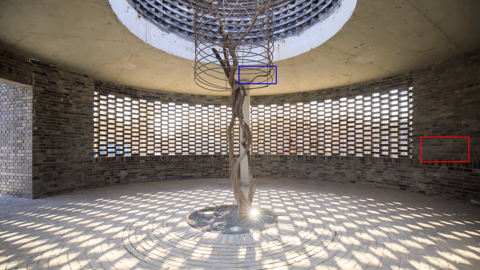}
	}
	\hspace{-3mm}
	\subfigure{
		\includegraphics[width=\gwidth]{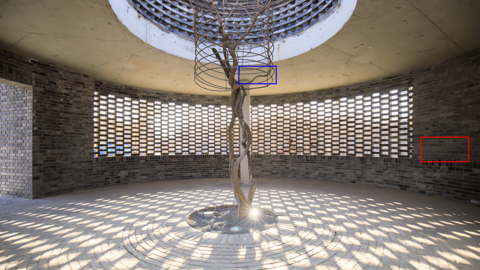}
	}
	\\ \vspace{-2mm}
	\subfigure{
		\includegraphics[width=\gwidth]{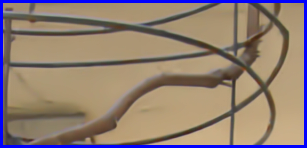}
	}
	\hspace{-3mm}
	\subfigure{
		\includegraphics[width=\gwidth]{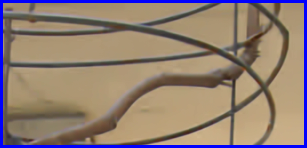}
	}
	\hspace{-3mm}
	\subfigure{
		\includegraphics[width=\gwidth]{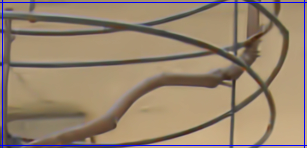}
	}
	\hspace{-3mm}
	\subfigure{
		\includegraphics[width=\gwidth]{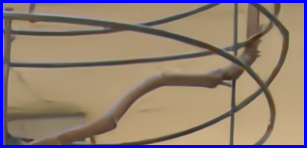}
	}
	\\ \vspace{-2mm}
	\vspace{-1mm}
	\setcounter{subfigure}{5}
	\subfigure[MWCNN~\cite{Liu_2018_CVPR_Workshops}]{
		\includegraphics[width=\gwidth]{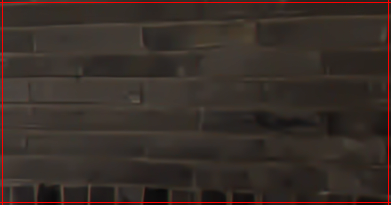}
	}
	\hspace{-3mm}
	\subfigure[MemNet~\cite{memnet}]{
		\includegraphics[width=\gwidth]{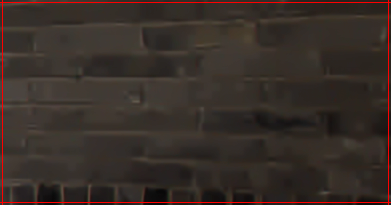}
	}
	\hspace{-3mm}
	\subfigure[PRN~\cite{PRN}]{
		\includegraphics[width=\gwidth]{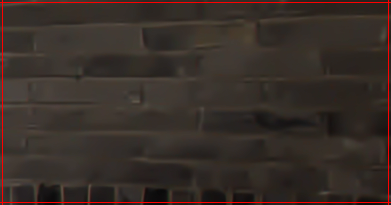}
	}
	\hspace{-3mm}
	\subfigure[DMCNN~\cite{DMCNN}]{
		\includegraphics[width=\gwidth]{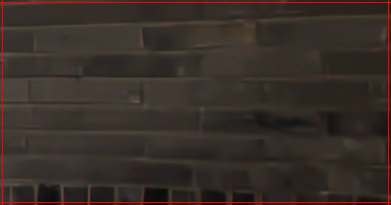}
	}
	\\
	\vspace{-1mm}	
	\caption{Examples of restored results on a compressed image by JPEG from LIU4K (QF=10).}
	\vspace{-5mm}	
	\label{fig:visual_comparison2}
\end{figure*}

\subsection{Evaluation Protocols}
Four full-reference metrics including PSNR, PSNR-B~\cite{psnr_b}, 
%$\Delta$PSNR, 
SSIM~\cite{ssim}, MS-SSIM~\cite{ms_ssim}, 
and two non-reference metrics including 
%SSEQ~\cite{LIU2014856}, 
NIQE~\cite{NIQE}, 
%BLINDS-II~\cite{BLIINDS_II},
and BRISQUE~\cite{BRISQUE}
are used to evaluate the effectiveness of the proposed method. 
In our implementation, we use Adam~\cite{Adam} optimizer to pre-train our network and finetune it with stochastic gradient descend (SGD)~\cite{Bottou10large} with cosine decay. 
In the first stage, the learning rate is set to 0.001. 
For PRN and MemNet, the learning rate is set to 0.0001.
After training 16 epochs, SGD is used for fintuning. The initial learning rate is set to 0.0001 at the second-stage training with cosine decay. We allow at most 60 epochs for JPEG artifacts removal and 300 epochs for restoration of compressed images by HEVC. 
For all methods, the models used for restorations of images compressed by JPEG with a quality factor 40 and HEVC with a quantization parameter 22 are trained from scratch. Other models are initialized these two models during the training.
\vspace{-3mm}

\subsection{Objective Comparison}

\wht{The objective results are presented in Table~\ref{tab:objective}.}
DMCNN is the obvious winner in full-reference metrics, 
followed by MWCNN for JPEG artifacts removal and PRN for loop filter. On the whole, deep learning-based methods achieve significantly superior performance than previous ones.
\wht{In no-reference metrics, TNRD achieves a superior performance for JPEG artifacts removal
and almost all methods generate results worse than original compressed images.}
\wht{We also provide more objective results of different methods on other testing sets in the supplementary material.} These results have high consensus levels among different testing sets.
\vspace{-3mm}

\begin{table*}[htbp]
	\footnotesize
	\centering
	\caption{Objective evaluations of different methods on LIU4K for compression artifacts \wht{reduction}. The value in {\color{red}red}, and {\color{blue}blue} colors denote the first, and \wh{second best} results, respectively.}
	\label{tab:objective}
	\vspace{-3mm}
	\begin{tabular}{cccccccccccc}
		\hline
		Method & Quality & Compressed                   & SA-DCT & TNRD   & ARCNN  & VRCNN  & DnCNN  & MemNet & MWCNN  & PRN    & DMCNN  \\
		\hline
		PSNR    & \multirow{6}{*}{QF=10} & 30.45      &  31.32  & 31.85  & 31.86  & 31.83  & 32.05  & 32.18  & {\color{blue}32.28}  & 32.19  & {\color{red}32.33}  \\
		PSNR-B  &                        & 29.93      &  31.32 & 31.82  & 31.83  & 31.81  & 32.01  & 32.10  & {\color{blue}32.23}  & 32.15  & {\color{red}32.30}  \\
		SSIM    &                        & 0.8090     & 0.8237 & 0.8427 & 0.8423 & 0.8419 & 0.8463 & 0.8488 & {\color{blue}0.8508} & 0.8491 & {\color{red}0.8520} \\
		MS-SSIM &                        & 0.9270     & 0.9353 & 0.9457 & 0.9457 & 0.9452 & 0.9481 & 0.9496 & {\color{blue}0.9511} & 0.9498 & {\color{red}0.9513} \\
		NIQE    &                        & 6.81       &  {\color{red}4.55}  & 5.31   & 5.22   & 5.31   & {\color{blue}5.16}   & 5.27   & 5.39   & 5.25   & 5.48   \\
		BRISQUE &                        & 56.39      & 51.09  & {\color{red}43.00}  & 49.94  & {\color{blue}48.63}  & 49.40  & 49.56  & 50.32  & 50.51  & 51.16  \\
		\hline
		PSNR    & \multirow{6}{*}{QF=20} & 33.28      &  33.87 & 34.48  & 34.51  & 34.47  & 34.57  & 34.80  & {\color{blue}34.86}  & 34.83  & {\color{red}34.91}  \\
		PSNR-B  &                        & 32.61      &  33.86 & 34.42  & 34.45  & 34.39  & 34.47  & 34.71  & {\color{blue}34.80}  & 34.78  & {\color{red}34.86}  \\
		SSIM    &                        & 0.8772     & 0.8787 & 0.8958 & 0.8963 & 0.8958 & 0.8970 & 0.9005 & {\color{blue}0.9013} & 0.9007 & {\color{red}0.9023} \\
		MS-SSIM &                        & 0.9675     & 0.9665 & 0.9737 & 0.9741 & 0.9738 & 0.9740 & 0.9757 & {\color{blue}0.9760} & 0.9757 & {\color{red}0.9762} \\
		NIQE    &                        & 5.30       &  {\color{red}4.23} & {\color{blue}4.84}   & {\color{blue}4.84}   & 4.98   & 4.86   & 4.94   & 4.96   & 4.94   & 5.17   \\
		BRISQUE &                        & 53.67      & 46.99 & {\color{red}39.44}  & 45.69  & 47.01  & {\color{blue}43.65}  & 46.62  & 45.85  & 46.20  & 47.10  \\
		\hline
		PSNR    & \multirow{6}{*}{QF=30} & 34.81      &  35.27 & 35.92  & 35.90  & 35.89  & 36.11  & 36.21  & {\color{blue}36.23}  & {\color{blue}36.23}  & {\color{red}36.31}  \\
		PSNR-B  &                        & 34.09      &  35.26 & 35.84  & 35.82  & 35.76  & 35.97  & 36.04  & {\color{blue}36.17}  & 36.16  & {\color{red}36.25}  \\
		SSIM    &                        & 0.9062     & 0.9040 & 0.9192 & 0.9196 & 0.9198 & 0.9215 & 0.9229 & 0.9232 & {\color{blue}0.9233} & {\color{red}0.9242} \\
		MS-SSIM &                        & 0.9799     & 0.9774 & 0.9830 & 0.9832 & 0.9833 & 0.9839 & 0.9842 & {\color{blue}0.9844} & 0.9843 & {\color{red}0.9846} \\
		NIQE    &                        & 4.68       &  {\color{red}4.08}  & {\color{blue}4.57}   & 4.64   & 4.59   & {\color{blue}4.57}   & 4.74   & 4.92   & 4.78   & 4.96   \\
		BRISQUE &                        & 48.48      &  45.28 & {\color{red}37.59}  & 43.10  & 42.64  & {\color{blue}42.49}  & 43.69  & 44.11  & 43.94  & 44.28  \\
		\hline
		PSNR    & \multirow{6}{*}{QF=40} & 35.82      &  36.20 & -      & 36.89  & 36.86  & 37.08  & 37.11  & 37.11  & {\color{blue}37.18}  & {\color{red}37.23}  \\
		PSNR-B  &                        & 35.07      & 36.19  & -      & 36.77  & 36.74  & 36.96  & 36.94  & 37.03  & {\color{blue}37.06}  & {\color{red}37.13}  \\
		SSIM    &                        & 0.9220     & 0.9188 & -      & 0.9331 & 0.9330 & 0.9349 & 0.9354 & 0.9353 & {\color{blue}0.9358} & {\color{red}0.9363} \\
		MS-SSIM &                        & 0.9856     & 0.9829 & -      & 0.9878 & 0.9878 & 0.9882 & 0.9883 & 0.9884 & {\color{blue}0.9885} & {\color{red}0.9886} \\
		NIQE    &                        & {\color{red}4.28}       & {\color{blue}4.00}   & -      & 4.49   & 4.54   & 4.63   & 4.61   & 4.75   & 4.68   & 4.80   \\
		BRISQUE &                        & 43.77      & 44.01  & -      & 41.35  & {\color{blue}41.00}  & {\color{red}40.79}  & 41.28  & 41.51  & 41.56  & 41.80  \\
		\hline
		PSNR    & \multirow{6}{*}{QP=22} & {\color{red}41.94}      & -      & -      & 41.72  & 41.72  & 41.79  & 41.80  & 41.79  & 41.75  & {\color{blue}41.86}  \\
		PSNR-B  &                        & 41.67      & -      & -      & 41.58  & 41.58  & {\color{blue}41.69}  & 41.67  & 41.68  & 41.62  & {\color{red}41.77}  \\
		SSIM    &                        & 0.9728     & -      & -      & 0.9729 & 0.9730 & {\color{blue}0.9732} & {\color{blue}0.9732} & {\color{blue}0.9732} & 0.9729 & {\color{red}0.9734} \\
		MS-SSIM &                        & {\color{blue}0.9957}     & -      & -      & {\color{blue}0.9957} & {\color{blue}0.9957} & {\color{blue}0.9957} & {\color{blue}0.9957} & {\color{blue}0.9957} & {\color{blue}0.9957} & {\color{red}0.9958} \\
		NIQE    &                        & 3.86       & -      & -      & {\color{blue}3.80}   & 3.81   & 3.93   & 3.96   & 3.91   & {\color{red}3.78}   & 3.97   \\
		BRISQUE &                        & {\color{red}21.29}      & -      & -      & 24.61  & 24.51  & 24.36  & 24.56  & 24.33  & {\color{blue}23.30}  & 24.57  \\
		\hline
		PSNR    & \multirow{6}{*}{QP=27} & 38.47      & -      & -      & 38.48  & 38.49  & 38.50  & 38.55  & 38.56  & {\color{red}38.61}  & {\color{blue}38.59}  \\
		PSNR-B  &                        & 38.33      & -      & -      & 38.43  & 38.45  & 38.47  & 38.50  & 38.52  & {\color{blue}38.56}  & {\color{red}38.57}  \\
		SSIM    &                        & 0.9456     & -      & -      & 0.9462 & 0.9462 & 0.9465 & 0.9465 & 0.9468 & {\color{red}0.9473} & {\color{blue}0.9472} \\
		MS-SSIM &                        & 0.9897     & -      & -      & 0.9896 & 0.9896 & {\color{blue}0.9898} & 0.9897 & {\color{blue}0.9898} & {\color{red}0.9899} & {\color{red}0.9899} \\
		NIQE    &                        & {\color{red}4.02}       & -      & -      & {\color{blue}4.06}   & 4.13   & 4.13   & 4.22   & 4.17   & 4.19   & 4.26   \\
		BRISQUE &                        & {\color{red}32.68}      & -      & -      & {\color{blue}34.65}  & 34.81  & 34.83  & 35.37  & 35.25  & 35.29  & 35.60  \\
		\hline
		PSNR    & \multirow{6}{*}{QP=32} & 35.52      & -      & -      & 35.63  & 35.65  & 35.71  & 35.74  & 35.75  & {\color{blue}35.78}  & {\color{red}35.79}  \\
		PSNR-B  &                        & 35.46      & -      & -      & 35.62  & 35.64  & 35.70  & 35.71  & 35.74  & {\color{blue}35.77}  & {\color{red}35.78}  \\
		SSIM    &                        & 0.9061     & -      & -      & 0.9072 & 0.9073 & 0.9083 & 0.9082 & 0.9085 & {\color{blue}0.9091} & {\color{red}0.9094} \\
		MS-SSIM &                        & 0.9776     & -      & -      & 0.9777 & 0.9776 & 0.9780 & 0.9779 & 0.9781 & {\color{blue}0.9782} & {\color{red}0.9783} \\
		NIQE    &                        & {\color{red}4.55}       & -      & -      & {\color{blue}4.57}   & 4.62   & 4.68   & 4.76   & 4.61   & 4.74   & 4.79   \\
		BRISQUE &                        & {\color{red}40.99}      & -      & -      & {\color{blue}41.87}  & 42.58  & 42.47  & 42.61  & 43.32  & 43.51  & 43.69  \\
		\hline
		PSNR    & \multirow{6}{*}{QP=37} & 32.85      & -      & -      & 33.00  & 33.02  & 33.06  & 33.12  & 33.14  & {\color{blue}33.16}  & {\color{red}33.17}  \\
		PSNR-B  &                        & 32.81      & -      & -      & 33.00  & 33.01  & 33.05  & 33.10  & {\color{blue}33.14}  & {\color{blue}33.14}  & {\color{red}33.16}  \\
		SSIM    &                        & 0.8558     & -      & -      & 0.8577 & 0.8582 & 0.8582 & 0.8594 & {\color{blue}0.8604} & 0.8603 & {\color{red}0.8613} \\
		MS-SSIM &                        & 0.9559     & -      & -      & 0.9563 & 0.9562 & 0.9564 & 0.9567 & {\color{blue}0.9573} & 0.9571 & {\color{red}0.9575} \\
		NIQE    &                        & {\color{blue}5.04}       & -      & -      & {\color{red}5.03}   & 5.07   & 5.07   & 5.24   & 5.08   & 5.14   & 5.21   \\
		BRISQUE &                        & {\color{red}46.61}      & -      & -      & 47.16  & 47.62  & {\color{blue}46.83}  & 48.74  & 47.77  & 47.11  & 48.30  \\
		\hline
	\end{tabular}
	\vspace{-1mm}
\end{table*}

\newcommand{\dred}[1]{{\color{red}#1}}
\newcommand{\dblue}[1]{{\color{blue}#1}}

\begin{table*}[t]
	\caption{The model complexity analysis of different methods. \textbf{J} denotes the version used for JPEG artifacts removal. \textbf{H} signifies the version used for the restoration of compressed images by HEVC.}
	\vspace{-2mm}
	\label{tab:complexity}
	\scriptsize
	\centering
	\begin{tabular}{cccccccccccc}
		\hline
		\multicolumn{1}{c}{Categories}     & \multicolumn{1}{c}{Non-Deep} & \multicolumn{7}{c}{Deep JPEG Artifacts Removal} & \multicolumn{3}{c}{Deep Loop Filter} \\
		\cmidrule(r){1-1} \cmidrule(r){2-2} \cmidrule(r){3-9} \cmidrule(r){10-12}
		Metrics                 & SA-DCT    & ARCNN    & TNRD      & DnCNN     & MemNet     & MWCNN       & PRN (J)     & DMCNN (J) & VRCNN  & PRN (H)   & DMCNN (H)   \\
		\hline
		Parameter               &   -       &  106,564 & 26,645    & 1,112,192 & 3,165,196  & 16,152,260  &  1,312,140  &   5,751,614  & 54,673 & 7,600,065 & 9,400,180  \\
		Storage (MB)            &   -       &   0.40   &  1.93     &   2.12    & 12.26      & 61.60       &  5.16       &  21.95       &  0.21  &  29.16    & 22.67      \\
		Time (ms/per-image)     & 43164.00  &   3.56   & 15050.30  &   6.31    & 186.30     & 132.39      &  49.92      &   17.34      &  3.92  &  841.01   & 32.48      \\
		\hline                
	\end{tabular}
	\vspace{-5mm}
\end{table*}

\subsection{Subjective Results}
We also compare the subjective quality of different methods in Fig.~\ref{fig:visual_comparison} and \ref{fig:visual_comparison2}. 
It is observed that, DMCNN achieves the overall best visual quality.
Most artifacts are removed and texture details are preserved due to its superior modeling capacity. 
As shown in Fig.~\ref{fig:visual_comparison}, JPEG, ARCNN, VRCNN and DnCNN generate obvious banding effects in large and smooth regions.
MemNet and PRN achieve better results. However, there are still gentle bands when taking a close look. Benefiting from the large receptive filed, MWCNN and DMCNN can well restore the artifacts in the smooth regions and remove banding artifacts. 
\wht{For the water wave textures, after compression, some regions are quantized to be small smooth blocks.
All methods fail to restore the visually pleasing texture. 
ARCNN, VRCNN, DnCNN only remove blockiness boundaries.
MemNet and PRN restore water wave textures in stochastic directions.
MWCNN and DMCNN generate consistent water wave textures to the surrounding waves.
Fig.~\ref{fig:visual_comparison2} provides the results of edges and regular textures. It is observed that, the results of ARCNN, VRCNN and DnCNN contain many artifacts. MWCNN, MemNet and PRN generate better results. DMCNN generates most shape edges and regular brick textures.}
\vspace{-5mm}

%As shown in Fig.~\ref{fig:visual_comparison3}, all methods will slightly smooth the texture details. More visual results will be presented in the supplementary material.

\wht{
	\begin{figure}[t]
		\footnotesize
		\centering
		\subfigure[]{
			\includegraphics[width=4.2cm]{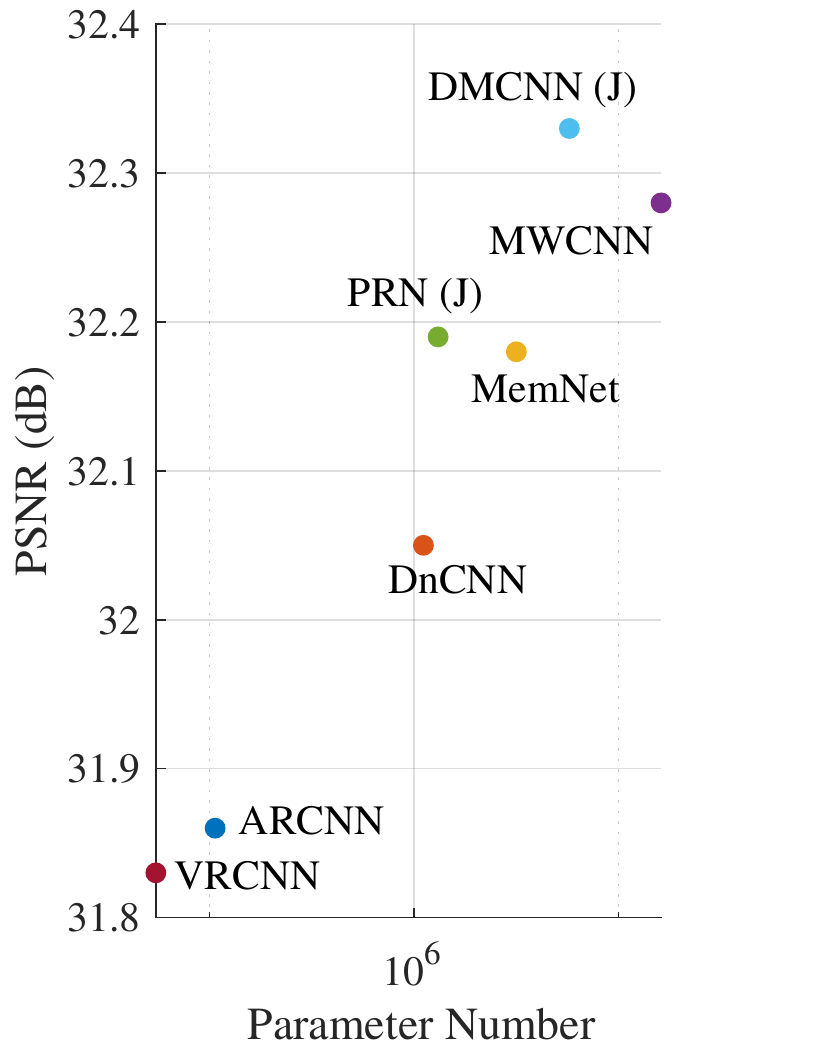}}
		\vspace{-3mm}
		\subfigure[]{
			\includegraphics[width=4.2cm]{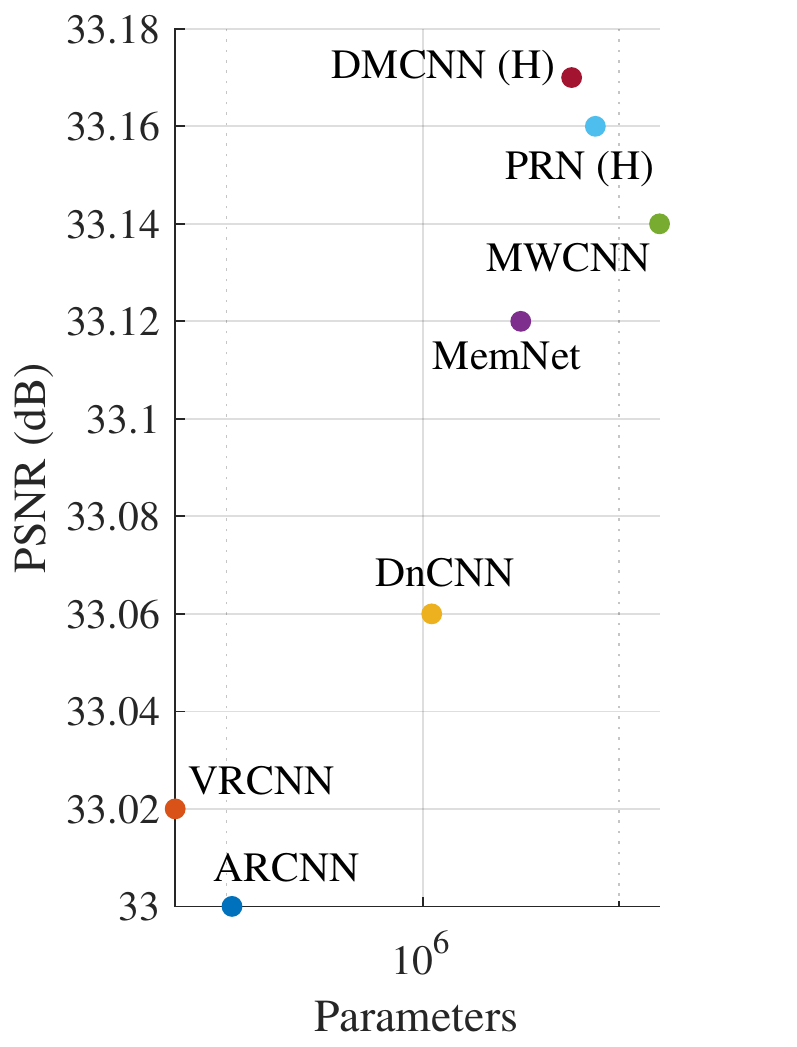}}	
		\caption{
			Visual results of performance and complexity (\textit{i.e.} parameters) of different methods. (a) JPEG artifacts removal (QF=10). (b) restoration of compressed images by HEVC (QP=37).
		}
		\label{fig:visual_complexity}
		\vspace{-2mm}
\end{figure}}

\subsection{Evaluation on Model Capacity}

Table~\ref{tab:complexity} reports the parameter number, the storage usage, and the per-image running time of each method, averaged over the images (768$\times$512) in LIVE1, 
\wht{on a machine with Intel(R) Xeon(TM) E5-2650 v4 2.20 GHz CPU, 16G RAM, and GeForce GTX 1080 Ti.}
ARCNN, DnCNN, MemNet, MWCNN, PRN, DMCNN, and VRCNN are implemented in Pytorch. 
SA-DCT and TNRD are implemented in MATLAB.
ARCNN, DnCNN, MemNet, MWCNN, PRN, DMCNN, and VRCNN run on GPU while SA-DCT and TNRD run on CPU.
It is observed that, all deep learning-based methods can finish process an image within 1 seconds. 
ARCNN, VRCNN and DnCNN achieve the shortest running time and can finish the restoration within 10 millisecond.
As for the storage, ARCNN and VRCNN use the minimum storage space.
As for the model complexity,  MWCNN uses the most parameters while TNRD uses the fewest ones.
Note that, PRN and DMCNN use different network architectures to handle JPEG artifacts removal and the restoration of compressed images by HEVC. Therefore, we present the complexities of their different versions in~Table~\ref{tab:complexity}, denoted with (J) and (H), respectively.
The results are also visualized in Fig.~\ref{fig:visual_complexity}.
\vspace{-3mm}

\subsection{Evaluations on Performance of Computer Vision Tasks}

\begin{figure}[t]
	\footnotesize
	\centering
	\subfigure{
		\includegraphics[width=2.1cm]{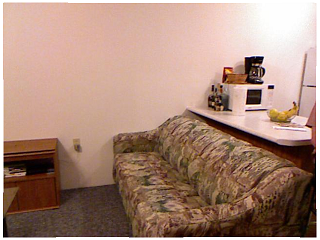}} \hspace{-2mm}
	\subfigure{
		\includegraphics[width=2.1cm]{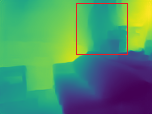}} \hspace{-2mm}
	\subfigure{
		\includegraphics[width=2.1cm]{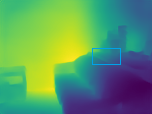}} \hspace{-2mm}
%	\subfigure{
%		\includegraphics[width=2.9cm]{./Images/Depth_estimation/00590_QF20_compressed.png}} \hspace{-2mm}
%	\subfigure{
%		\includegraphics[width=2.9cm]{./Images/Depth_estimation/00590_QF20_predicted.png}} \hspace{-2mm}
	\subfigure{
		\includegraphics[width=2.1cm]{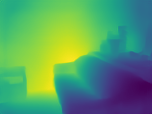}} 
	\\ \vspace{-3mm} \setcounter{subfigure}{0}
	\subfigure[]{
		\includegraphics[width=2.1cm]{./Images/Depth_estimation/01010_colors.png}} \hspace{-2mm}
	\subfigure[]{
		\includegraphics[width=2.1cm]{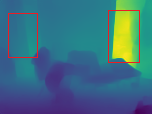}} \hspace{-2mm}
	\subfigure[]{
		\includegraphics[width=2.1cm]{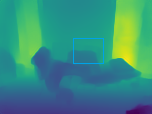}}
		\hspace{-2mm}
%	\subfigure[]{
%		\includegraphics[width=2.9cm]{./Images/Depth_estimation/01010_QF20_compressed.png}} \hspace{-2mm}
%	\subfigure[]{
%		\includegraphics[width=2.9cm]{./Images/Depth_estimation/01010_QF20_predicted.png}} \hspace{-2mm}	
	\subfigure[]{
		\includegraphics[width=2.1cm]{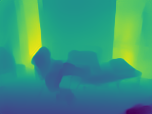}}
	\\
	\vspace{-2mm}
	\caption{
	The visual results of depth estimation on compressed images (JPEG) with and without compression artifacts \wht{reduction}. 
	(a) Input RGB image. (b) Depth map of compressed image (QF=10). (c)  Depth map of restored image (QF=10). (d) Depth map of original image. 
%	(e) Depth map of compressed image (QF=20). 
%	(f) Depth map of restored image (QF=20).
	}\vspace{-5mm}
	\label{fig:visual_result}
\end{figure}

\wht{
	\begin{figure}[t]
		\footnotesize
		\centering
		\subfigure[]{
			\includegraphics[width=4.2cm]{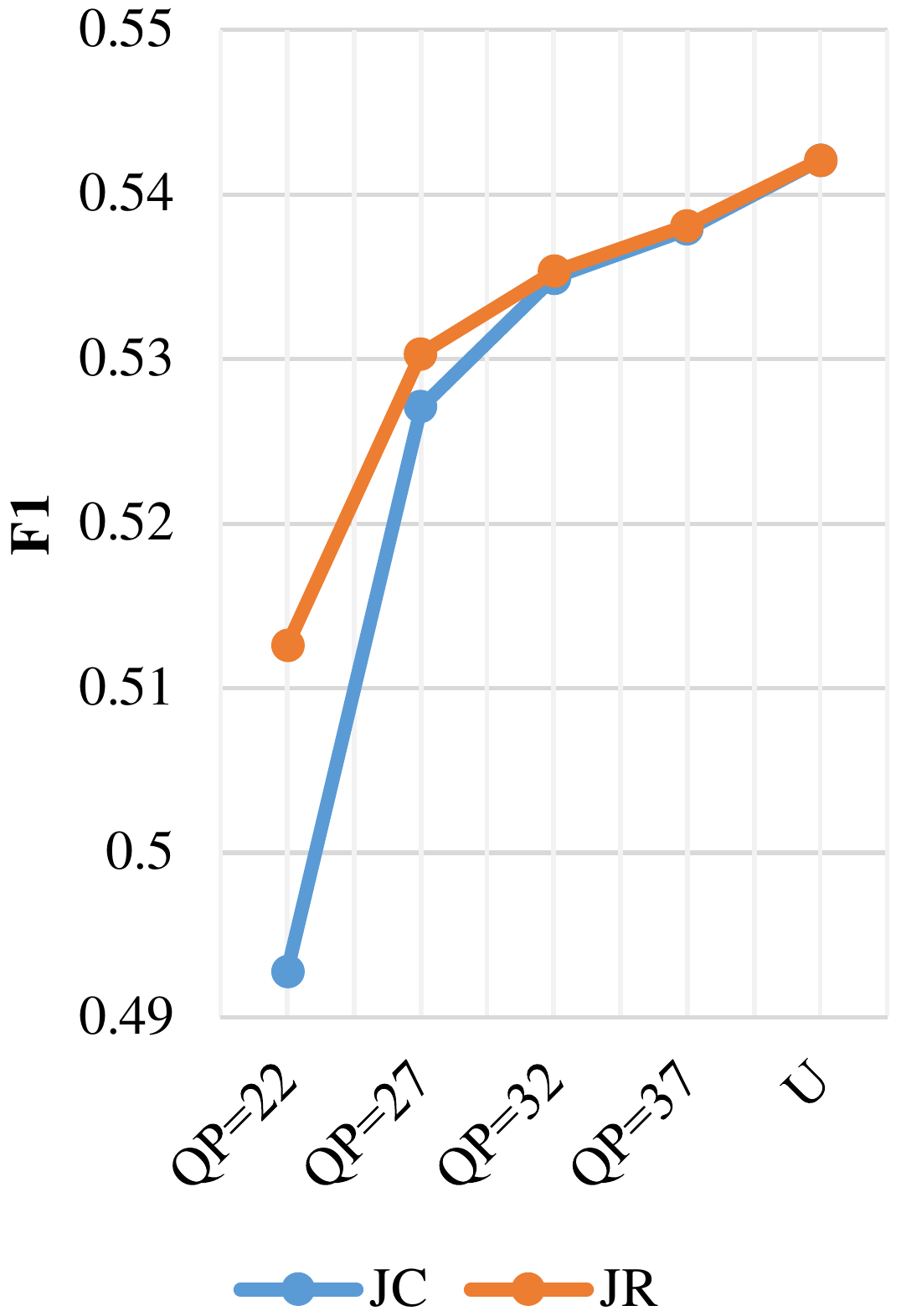}} \hspace{-1mm}
		\subfigure[]{
			\includegraphics[width=4.2cm]{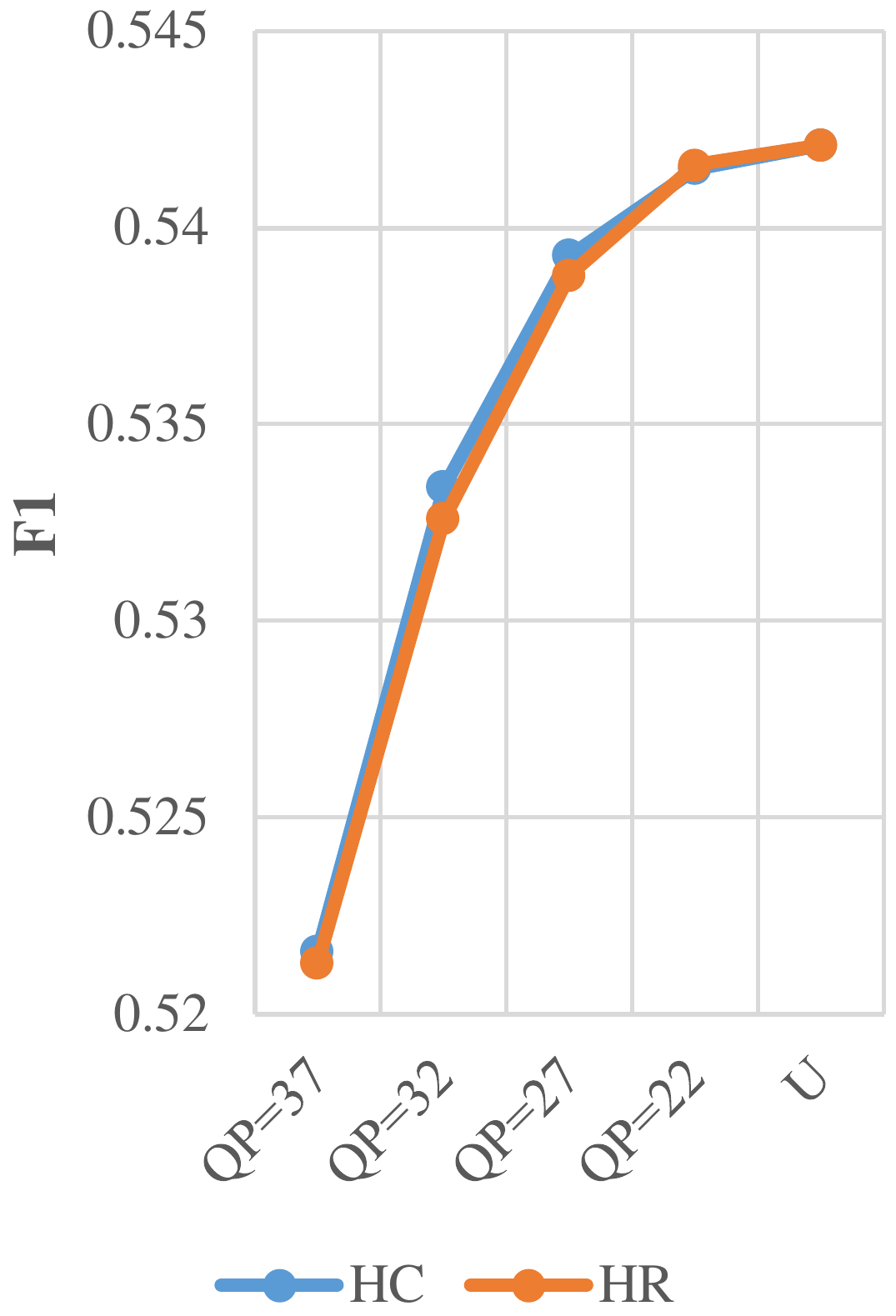}} \vspace{-2mm}
		\caption{
			Visual results of performance changes before and after the restoration at different QP/QF conditions for depth estimation.
			JC: Compressed by JPEG.
			HC: Compressed by HEVC.
			JR: Restored from images compressed by JPEG.
			HR: Restored from images compressed images by HEVC.
		}
		\label{fig:accuracy_depth}
		\vspace{-4mm}
	\end{figure}
}

\begin{figure}[t]
	\footnotesize
	\centering
	\subfigure{
		\includegraphics[width=2.1cm]{./Images/Segmentation/Original.png}} \hspace{-2mm}
	\subfigure{
		\includegraphics[width=2.1cm]{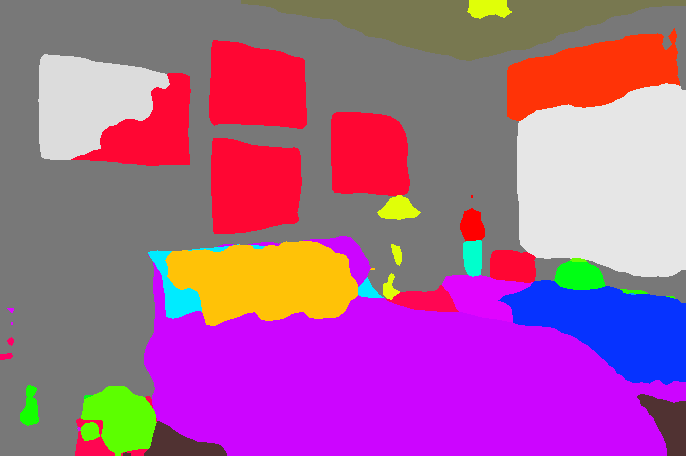}} \hspace{-2mm}
	\subfigure{
		\includegraphics[width=2.1cm]{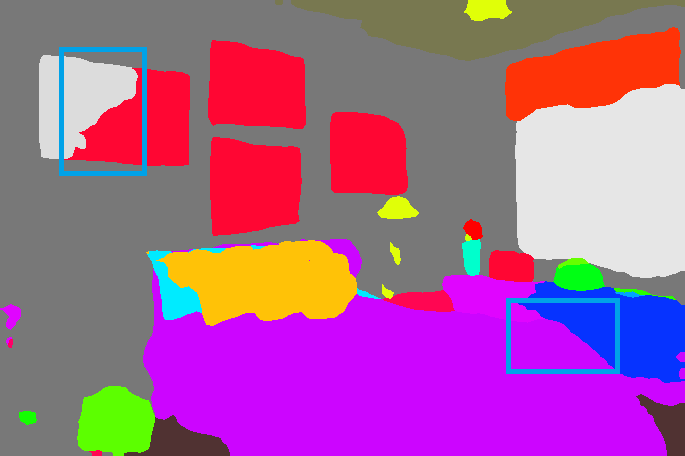}} \hspace{-2mm}
	\subfigure{
		\includegraphics[width=2.1cm]{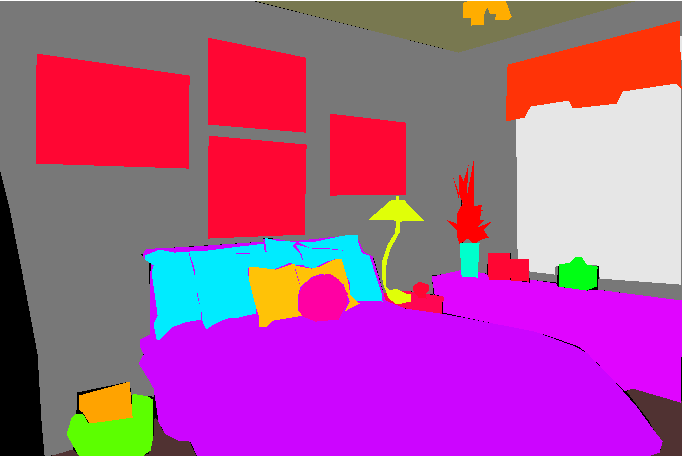}} 
	\\ 
	\vspace{-2mm} \setcounter{subfigure}{0}
	\subfigure[]{
		\includegraphics[width=2.1cm]{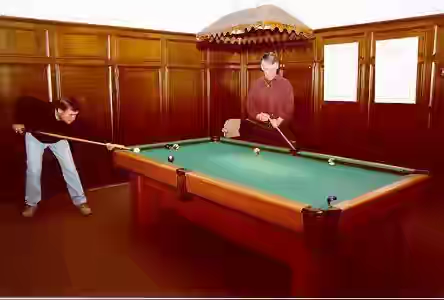}} \hspace{-2mm}
	\subfigure[]{
		\includegraphics[width=2.1cm]{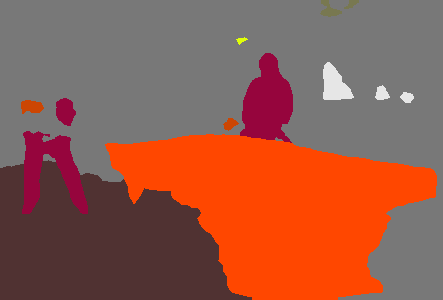}} \hspace{-2mm}
	\subfigure[]{
		\includegraphics[width=2.1cm]{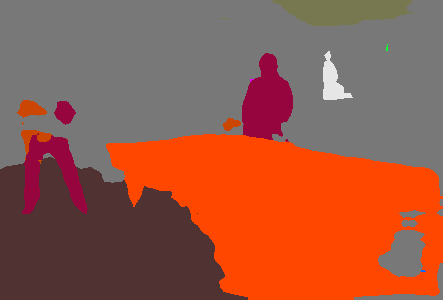}}
	\hspace{-2mm}	
	\subfigure[]{
		\includegraphics[width=2.1cm]{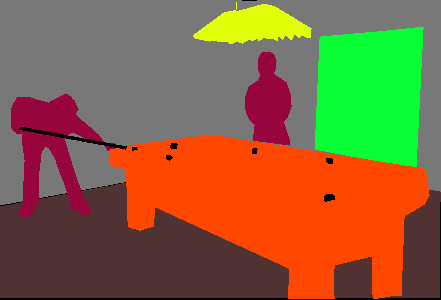}}
	\\ 
	\vspace{-2mm}
	\caption{
		The visual results of semantic segmentation on compressed images (HEVC) with and without compression artifacts \wht{reduction}. 
		(a) Input RGB image. (b) Semantic map of compressed image (QP=37). (c)  Semantic map of restored image (QP=37). (d) Semantic map of original image. 
	}
	\vspace{-6mm}
	\label{fig:visual_result_seg}
\end{figure}
\wht{
	\begin{figure}[t]
		\footnotesize
		\centering
		\subfigure[]{
			\includegraphics[width=4.2cm]{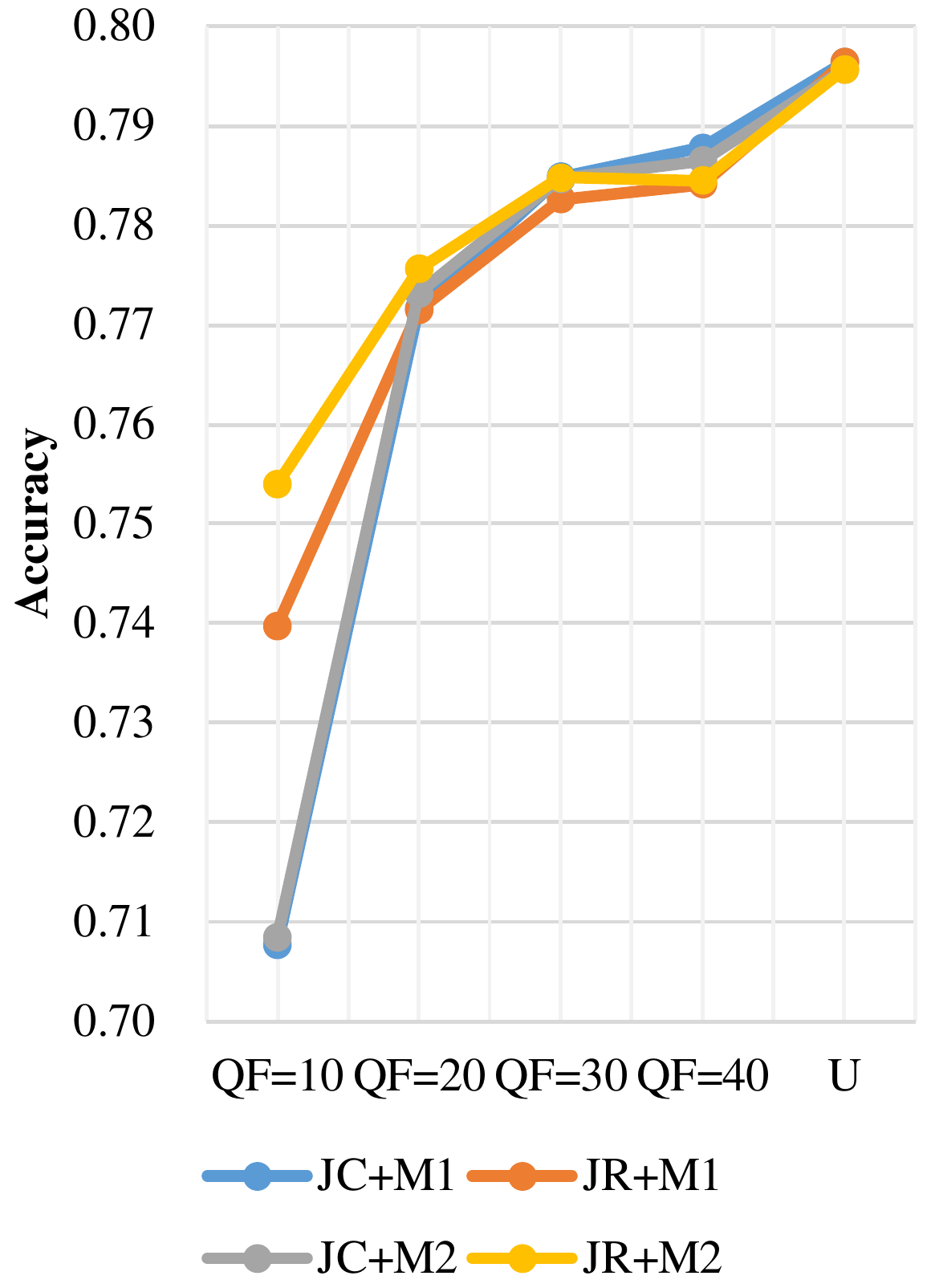}} \hspace{-1mm}
		\subfigure[]{
			\includegraphics[width=4.2cm]{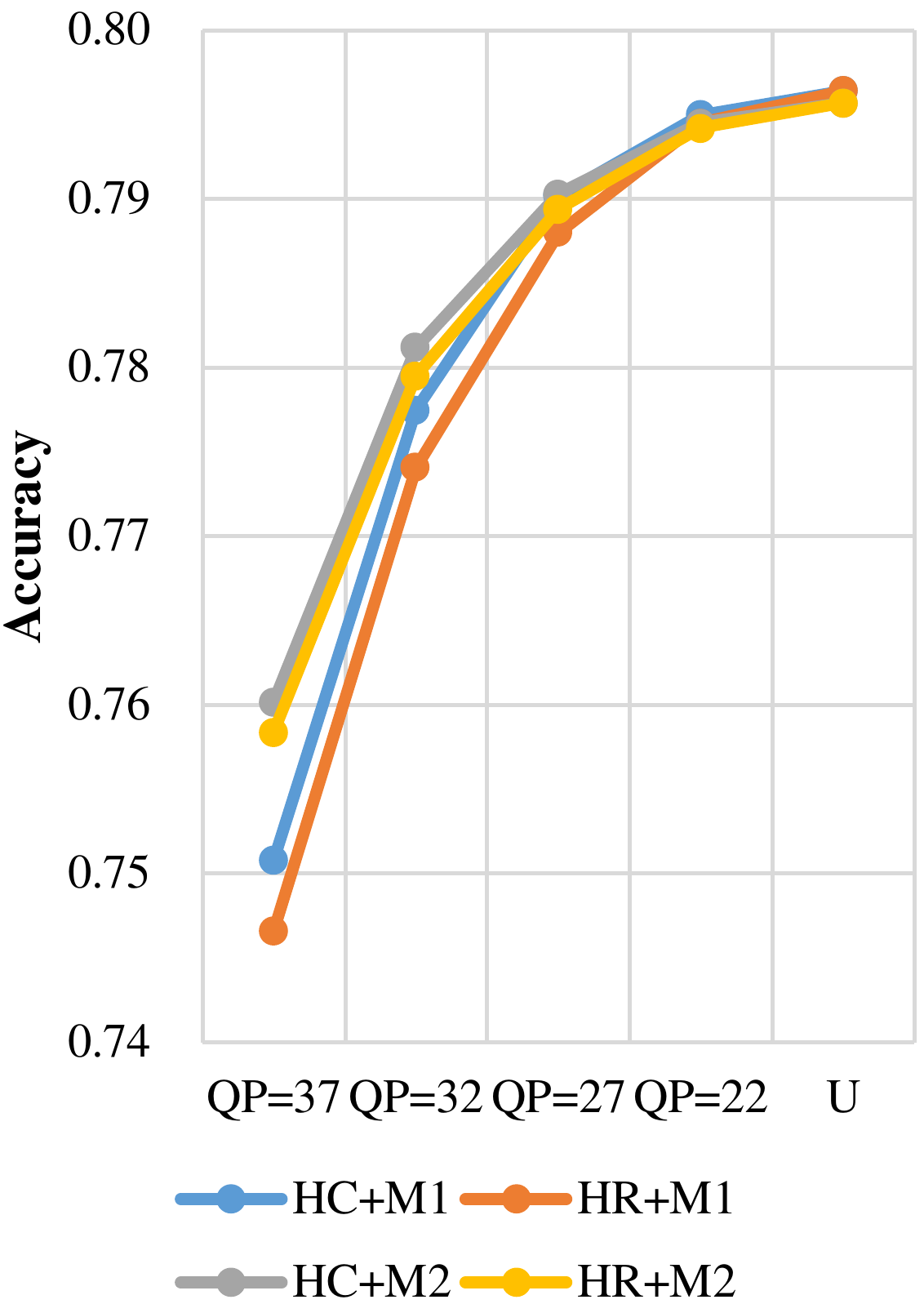}}	\vspace{-2mm}
		\caption{
			Visual results of performance changes before and after the restoration at different QP/QF conditions for semantic segmentation.
			JC: Compressed by JPEG.
			HC: Compressed by HEVC.
			JR: Restored from images compressed by JPEG.
			HR: Restored from images compressed images by HEVC.
			M1: ResNet50Dilated + PPM Deepsup.
			M2: ResNet50 + UperNet.
		}
		\label{fig:accuracy_semantic}
		\vspace{-6mm}
\end{figure}}

\begin{table*}[]
	\footnotesize
	\centering
	\caption{
		Comparisons of SENet on the NYU-Depth V2 dataset. 
		The input testing images in ``Compressed'' category are compressed with different quantization parameters.
		The input testing images in ``Restored'' category are processed with VRCNN.
		``QF'' denotes the quality factor.
		``QP'' signifies the quantization parameter.
	}\vspace{-3mm}
	\label{}
	\begin{tabular}{ccccccccccc}
		\hline
		Metrics           & Codecs                                   & Uncompressed & \multicolumn{4}{c}{Compressed}    & \multicolumn{4}{c}{Restored}     \\
		\cmidrule(r){1-1} \cmidrule(r){2-2} \cmidrule(r){3-3} \cmidrule(r){4-7}  \cmidrule(r){8-11} 
		QF                & -                                        & -            & QF=10  & QF=20  & QF=30  & QF=40  & QF=10  & QF=20  & QF=30  & QF=40  \\
		\hline
		MSE               &                                          & 0.2994       & 0.4072 & 0.3193 & 0.3080 & 0.3031 & 0.3652 & 0.3163 & 0.3058 & 0.3021  \\
		RMS               &          \multirow{9}{*}{JPEG}           & 0.5471       & 0.6382 & 0.5651 & 0.5550 & 0.5505 & 0.6043 & 0.5624 & 0.5530 & 0.5496 \\
		REL               &                                          & 0.1179       & 0.1482 & 0.1248 & 0.1207 & 0.1191 & 0.1354 & 0.1245 & 0.1214 & 0.1201 \\
		log10             &                                          & 0.0521       & 0.0644 & 0.0548 & 0.0531 & 0.0526 & 0.0611 & 0.0551 & 0.0535 & 0.0530 \\
		%		MAE               &                                          & 0.3260       & 0.3710 & 0.3396 & 0.3319 & 0.3294 & 0.3951 & 0.3397 & 0.3315 & 0.3286 \\
		$\delta<1.25$     &                                          & 0.8572       & 0.7910 & 0.8414 & 0.8529 & 0.8546 & 0.8057 & 0.8384 & 0.8488 & 0.8520 \\
		$\delta<1.25$     &                                          & 0.9728       & 0.9484 & 0.9689 & 0.9711 & 0.9720 & 0.9531 & 0.9679 & 0.9703 & 0.9715 \\
		$\delta<1.25^3$   &                                          & 0.9928       & 0.9864 & 0.9914 & 0.9921 & 0.9927 & 0.9888 & 0.9917 & 0.9922 & 0.9926 \\
		P                 &                                          & 0.6220       & 0.5694 & 0.6091 & 0.6167 & 0.6193 & 0.6020 & 0.6099 & 0.6133 & 0.6156 \\
		R                 &                                          & 0.4904       & 0.4481 & 0.4751 & 0.4825 & 0.4853 & 0.4575 & 0.4795 & 0.4851 & 0.4880 \\
		F1                &                                          & 0.5421       & 0.4928 & 0.5271 & 0.5349 & 0.5379 & 0.5126 & 0.5303 & 0.5353 & 0.5381 \\		
		\hline
		QP & - & - & QP=22 & QP=27 & QP=32 & QP=37 & QP=22 & QP=27 & QP=32 & QP=37  \\
		\hline
		MSE               &                                            &  0.2994  & 0.2976 & 0.2973 & 0.3129 & 0.3385 & 0.2971 & 0.2991 & 0.3183 & 0.3483 \\
		RMS               & \multicolumn{1}{c}{\multirow{9}{*}{HEVC}} & 0.5471 & 0.5455 & 0.5453 & 0.5594 & 0.5818 & 0.5451 & 0.5469 & 0.5642 & 0.5901 \\
		REL               &                                            & 0.1179 & 0.1185 & 0.1199 & 0.1235 & 0.1300 & 0.1188 & 0.1204 & 0.1246 & 0.1319 \\
		log10             &                                            & 0.0521 & 0.0522 & 0.0527 & 0.0546 & 0.0580 & 0.0523 & 0.0530 & 0.0553 & 0.0594 \\
		$\delta<1.25$     &                                            & 0.8572 & 0.8553 & 0.8535 & 0.8420 & 0.8247 & 0.8547 & 0.8517 & 0.8376 & 0.8171 \\
		$\delta<1.25^2$   &                                            & 0.9728 & 0.9725 & 0.9714 & 0.9682 & 0.9617 & 0.9724 & 0.9709 & 0.9664 & 0.9581 \\
		$\delta<1.25^3$   &                                            & 0.9928 & 0.9929 & 0.9926 & 0.9918 & 0.9900 & 0.9930 & 0.9924 & 0.9916 & 0.9895 \\
		P                 &                                            & 0.6220 & 0.6196 & 0.6133 & 0.6090 & 0.6030 & 0.6188 & 0.6130 & 0.6098 & 0.6065 \\
		R                 &                                            & 0.4904 & 0.4909 & 0.4913 & 0.4847 & 0.4710 & 0.4915 & 0.4909 & 0.4833 & 0.4687 \\
		F1                &                                            & 0.5421 & 0.5415 & 0.5393 & 0.5334 & 0.5216 & 0.5416 & 0.5388 & 0.5326 & 0.5213 \\
		\hline
	\end{tabular}
	\vspace{-2mm}
\end{table*}

\begin{table*}[t]
	\centering
	\scriptsize
	\caption{
		Comparisons of SENet on the NYU-Depth V2 dataset. 
		The input testing images in ``Compressed'' category are compressed with different quantization parameters.
		The input testing images in ``Predicted'' category are processed with VRCNN.
		``QF'' denotes the quality factor.
		``QP'' signifies the quantization parameter.
	}\vspace{-3mm}
	\label{tab:cv_task_objective}
	\begin{tabular}{cccccccccccc}
		\hline
		Metrics  & Codecs                & Baseline                                        & Uncompressed & \multicolumn{4}{c}{Compressed}        & \multicolumn{4}{c}{Restored}         \\
		\cmidrule(r){1-1}  \cmidrule(r){2-2} \cmidrule(r){3-3} \cmidrule(r){4-4} \cmidrule(r){5-8} \cmidrule(r){9-12}
		QF       & -                     & -                                               & -            & QF=10   & QF=20   & QF=30   & QF=40   & QF=10   & QF=20   & QF=30   & QF=40   \\
		
		Mean IoU & \multirow{4}{*}{JPEG} & \multirow{2}{*}{ResNet50Dilated + PPM\_Deepsup}   & 0.4075       & 0.2794  & 0.3711  & 0.3911  & 0.394   & 0.3145  & 0.3624  & 0.382   & 0.3831  \\
		Accuracy &                       &                                                   & 79.64\%      & 70.77\% & 77.17\% & 78.49\% & 78.78\% & 73.98\% & 77.15\% & 78.26\% & 78.41\% \\
		Mean IoU &                       & \multirow{2}{*}{ResNet50 + UperNet}               & 0.4029       & 0.2814  & 0.3662  & 0.3842  & 0.3884  & 0.3345  & 0.3685  & 0.3834  & 0.3843  \\
		Accuracy &                       &                                                   & 79.57\%      & 70.85\% & 77.32\% & 78.47\% & 78.65\% & 75.40\% & 77.56\% & 78.48\% & 78.45\% \\
		\hline
		QP       & -                     & -                                                 & -            & QP=22   & QP=27   & QP=32   & QP=37   & QP=22   & QP=27   & QP=32   & QP=37   \\		
		Mean IoU & \multirow{4}{*}{HEVC} & \multirow{2}{*}{ResNet50Dilated + PPM\_Deepsup}   &  0.4075      & 0.4043  &  0.3956 &  0.3736 & 0.3363  & 0.4027  & 0.3902  & 0.3653  &  0.3269	\\
		Accuracy &                       &                                                   &  79.64\%     & 79.50\% & 79.02\% & 77.75\% & 75.08\% & 79.45\% & 78.80\% & 77.41\% & 74.66\% \\
		Mean IoU &                       & \multirow{2}{*}{ResNet50 + UperNet}               &  0.4029      & 0.4006  & 0.3945  & 0.3789  & 0.3463  & 0.3999  & 0.3931  & 0.3744  & 0.3417  \\ 
		Accuracy &                       &                                                   &  79.57\%     & 79.45\% & 79.03\% & 78.12\% & 76.02\% & 79.42\% & 78.94\% & 77.95\% & 75.84\% \\
		\hline
	\end{tabular}
	\vspace{-2mm}
\end{table*}

\noindent \textbf{Depth Estimation}. Table~\ref{tab:cv_task_objective} shows the results by depth estimation with accurate object boundaries~\cite{Hu2018RevisitingSI}, one of the state-of-the-art depth estimation methods, based on images with and without compression artifacts \wht{reduction} by VRCNN on NYUv2~\cite{nyuv2} in different measures. 
Several accuracy measures are employed to evaluate the depth estimation performance. Mean squared error (MSE), root mean squared error (RMS), mean relative error (MRE), mean log 10 error (log 10), and threshold accuracy, as well as the precision (P), recall (R) and F1 score of the estimated edge maps.
It is noted that, for MSE, RMS and MRE, small values signify better performance.
For log 10, threshold accuracy $\delta$, P, R and F1 score, large values denote better performance.
From the results, for MSE, RMS, and MRE, it is always beneficial to perform compression artifacts \wht{reduction} among all cases (both JPEG and HEVC codecs with all QPs and QFs). For other metrics, the results become slightly controversial. 
The results of the restored images are sometimes inferior to those of the compressed ones, \textit{e.g.}, QF=10 on log 10, and QF=30, 40 on P, \textit{etc}. However, in general, the results of restored images win in more cases, compared to the compressed ones. It is also demonstrated that, the reconstruction aiming to restore compressed images in visual quality might be not beneficial to the successive tasks all the time.
\wht{The trend of the performance change before and after the restoration at different QP/QF conditions is also visualized in Fig.~\ref{fig:accuracy_depth}.}
We also show the visual results in Fig.~\ref{fig:visual_result}. 
It is observed that, when QF=10, the result of the compressed image degrades much and the enhancement operation effectively improves the visual quality of the depth map. When QF=20, the degradations in the result of the compressed image are not obvious. The enhancement operation also leads to minor visual quality gains.
Some discontinuous boundary artifacts are removed as shown in the red boxes of Fig~\ref{fig:visual_result}. 
However, some details become blurry, \textit{e.g.} the details and boundaries in the blue boxes, as shown in Fig~\ref{fig:visual_result}.

\noindent \textbf{Semantic Segmentation}. We integrate two baselines for evaluations: ResNet50Dilated + PPM\_Deepsup and ResNet50 + UperNet~\cite{zhou2017scene}. The evaluation is performed on ADE20K~\cite{zhou2017scene}. Results are reported in two metrics commonly used for semantic segmentation~\cite{FCN}:
{Pixel accuracy} indicates the proportion of correctly classified pixels. 
{Mean IoU} indicates the intersection-over-union between the predicted and groundtruth pixels, averaged over all the classes.
It is observed from Table~\ref{tab:cv_task_objective} that, compression artifacts \wht{reduction} (\textit{i.e.} VRCNN) may not benefit the inference of semantic segmentation all the time. In many cases, \textit{e.g.} for JPEG artifacts, the performance of the baseline ResNet50Dilated + PPM\_Deepsup on restored images is worse than that on compressed images for QF=40.
\wht{The trend of the performance change before and after the restoration at different QP/QF conditions is also visualized in Fig.~\ref{fig:accuracy_semantic}.}
\wht{The main reason of the performance drop might be the consensus of the effect of MSE used in the training and the semantic segmentation purpose. Training with MSE, the restoration results of compressed images with gender artifacts tend to be smooth and some critical details are lost causing low accuracy.}
For visual results, it is observed from Fig.~\ref{fig:visual_result_seg} that, compression artifact removal might \wh{slightly} correct some false boundaries.
\vspace{-1mm}

\section{Trends and Challenges}
\vspace{-1mm}

Although deep learning techniques bring fast development in compression artifacts \wht{reduction},
there remains several important challenges and inherent trends. 
First, recent researchers obtained higher and higher accuracy by advanced deep models with a huge amount of parameters, it is still hard to apply these methods in real scenario. It is interesting to re-design compact deep network architectures and compress or adjust the existing models into small ones for \textit{real-time compression artifacts \wht{reduction}}.
Second, with the latest codecs, \textit{i.e.} versatile video coding (VVC), more integrated tools are employed, thus the distribution of compression artifacts is more complex. It is  challenging to apply the existing methods to the next generation codecs. With more powerful weapons of deep learning, \textit{e.g.} capsule network, and reinforcement learning \textit{etc.}, we believe that, the future \textit{technique improvement on restoration of more complex degradations} will bring new surprises. Third, for compression artifacts \wht{reduction}, there are few works on the \textit{internal mechanism of feature learning and the related interpretable factors}. Beyond obtaining superior performance, one direction is to give comprehensive explanations on what factors might lead to a more effective network and the specific mechanism. Last, for various low-level image processing tasks, it is a critical issue to design and apply proper metrics to constrain model training and evaluate the model's effectiveness. Thus, it is an important future direction to develop more effective and rational measures that balance both signal fidelity and visual perception for compression artifacts \wht{reduction}.

\section{Conclusion}
\label{sec:conclusion}

This paper presents a systematic review of compression artifacts \wht{reduction} methods, including both traditional and deep-learning based methods.
These methods evolve in several perspectives, including model architecture improvement and continuing exploration of side information embedding, \textit{etc}.
We summarize milestone and typical methods and highlight their contributions, strengths, and weaknesses.
We also conduct a thorough benchmark of state-of-the-art compression artifacts \wht{reduction} methods.
In our benchmarking experiments, some constraints and training skills targeted for JPEG artifacts removal are generalized to handle general compression artifacts \wht{reduction} methods.
Based on our evaluation and analysis, overall remarks, challenges, and trends are given.
Although our attempts are preliminary, they build a bridge from the existing world to a new one, where more researchers are expected to come.
\vspace{-3mm}

\renewcommand{\baselinestretch}{0.9}

%
%\bibliographystyle{IEEEbib}
%%\bibliographystyle{plain}
%
%
%\bibliography{sigproc_v2}

\end{document}